# Nucleosome positioning and energetics: Recent advances in genomic and computational studies

Denis Tolkunov and Alexandre V. Morozov [1]

Department of Physics & Astronomy and BioMaPS Institute for Quantitative Biology, Rutgers University, Piscataway, NJ 08854

[1] Corresponding author: _morozov@physics.rutgers.edu_, phone: 732-445-1387, fax: 732-445-5958

I. Introduction

II. Experimental studies of chromatin structure

      1. Genome-wide mapping of nucleosome positions

      2. Structural studies of the nucleosome core particle

III. Computational studies of chromatin structure

      1. Using DNA elasticity theory to predict nucleosome formation energies

      2. Bioinformatics models of nucleosome sequence preferences

      3. Statistical physics of one-dimensional liquids and the nucleosome positioning problem

      4. Hidden Markov Models for predicting nucleosome occupancies

IV. Summary and Conclusions

References

## Abstract

Chromatin is a complex of DNA, RNA and proteins whose primary function is to package genomic DNA into the tight confines of a cell nucleus. A fundamental repeating unit of chromatin is the nucleosome, an octamer of histone proteins around which 147 base pairs of DNA are wound in almost two turns of a left-handed superhelix. Chromatin is a dynamic structure which exerts profound influence on regulation of gene expression and other cellular functions. These chromatin-directed processes are facilitated by optimizing nucleosome positions throughout the genome and by remodeling nucleosomes in response to various external and internal signals such as environmental perturbations. Here we discuss large-scale maps of nucleosome positions made available through recent advances in parallel high-throughput sequencing and microarray technologies. We show that these maps reveal common features of nucleosome organization in eukaryotic genomes. We also survey computational models designed to predict nucleosome formation scores or energies, and demonstrate how these predictions can be used to position multiple nucleosome on the genome without steric overlap.

# I. Introduction

DNA in eukaryotic nuclei is assembled into chromatin – a complex combination of DNA, RNA and proteins that makes up chromosomes. The primary function of chromatin is to compact genomic DNA which otherwise would not fit into the cell nucleus. However, since the early days of chromatin studies (Kornberg and Thomas 1974) it has been recognized that chromatin function goes well beyond DNA compaction: in particular, chromatin exerts a profound influence on gene regulation, replication, recombination, and DNA repair by both blocking access to DNA (Boeger *et al.* 2003) and by juxtaposing sites far apart on the linear sequence (Wallrath *et al.* 1994).

The building block of chromatin is the nucleosome core particle (Khorasanizadeh 2004) , a 147 base pair (bp) long DNA segment wrapped in ~1.65 superhelical turns around the surface of a histone octamer (Luger *et al.* 1997; Richmond and Davey 2003) (Figure 1a). On the first level of compaction, DNA is arranged into one-dimensional quasi-periodic nucleosomal arrays,

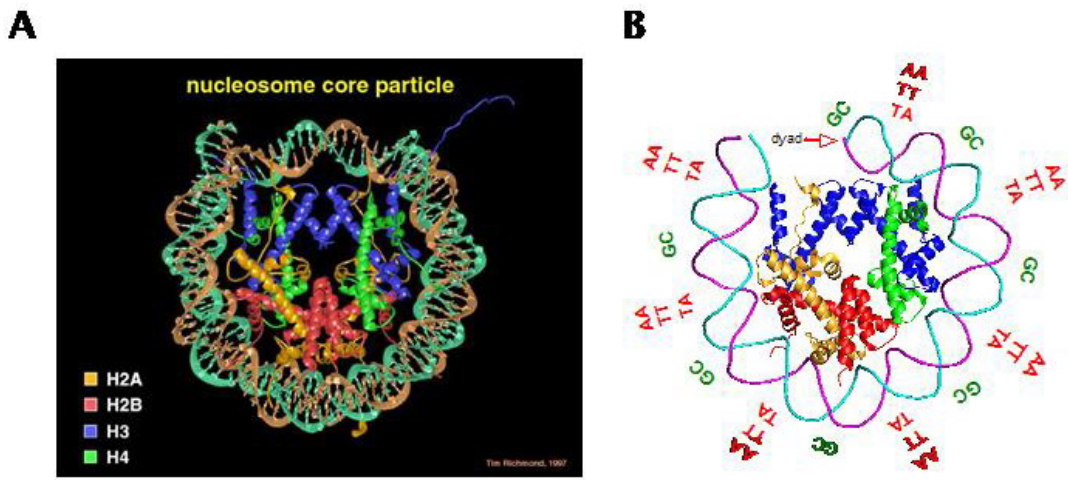

**Figure 1** a) Crystal structure of the nucleosome core particle (Luger *et al.* 1997) (courtesy of Song Tan, Penn State University). 147 bp long DNA is wrapped around the histone octamer in ~1.65 turns of a left-handed superhelix. The histone octamer consists of two copies each of histones H2A (yellow), H2B (red), H3 (blue) and H4 (green). b) Bending of nucleosomal DNA is mediated by specific dinucleotides located at positions where DNA minor or major groove faces the histone octamer (reproduced with permission from (Segal *et al.* 2006)). Relative frequencies of A/T-rich dinucleotides tend to increase at positions where the minor groove faces the surface of the histone octamer, whereas relative frequencies of G/C-rich dinucleotides tend to increase where the minor groove faces away and the major groove faces toward the histone octamer.

which in turn fold into higher-order chromatin fibers (Felsenfeld and Groudine 2003). Chromatin fiber formation is stabilized in part by the linker histone H1. Neighboring nucleosomes (which consist of the nucleosome core particle and H1) are separated from each other by ~10–60 bp stretches of linker DNA, which means that ~70–90% of genomic DNA is wrapped in nucleosomes. The histone octamer is made of two copies of four highly conserved histone proteins: H2A, H2B, H3 and H4. Histones have unstructured tail domains that protrude from the surface of the histone octamer, providing sites for potential interactions with other proteins. Histone tails are targets of numerous post-translational covalent modifications such as acetylation, phosphorylation, methylation, ubiquitination and ADP-ribosylation, and may also influence how nucleosome arrays fold into higher-order chromatin structures (Strahl and Allis 2000; Khorasanizadeh 2004).

In this review we focus on genome-wide predictions of nucleosome formation energies and positions obtained by analyzing DNA conformational properties and high-throughput nucleosome mapping data. DNA sequences differ greatly in their ability to form nucleosomes - *in vitro* studies show that the range of histone-DNA binding affinities is at least a thousand-fold (Thastrom *et al.* 1999). Nucleosomal DNA is sharply bent to achieve tight wrapping around the histone octamer. This bending occurs at every 10-11 bp DNA helical repeat, when the minor groove of the DNA faces inwards toward the histone octamer, and again ~5 bp away, with opposite direction, when the minor groove faces outwards (Figure 1b). Bends of each direction are facilitated by specific dinucleotides – up to higher-order effects, sequence-specific DNA bending is controlled by base stacking energies between neighboring base pairs.

It is reasonable to assume that *in vitro* nucleosome positions are determined purely by intrinsic sequence preferences and by steric exclusion between neighboring nucleosomes. *In vivo* however nucleosomes compete with non-histone DNA-binding factors for access to genomic DNA, which may result in overriding intrinsic sequence preferences. In addition, chromatin remodeling enzymes play a role that needs to be quantified: in one scenario the role of such enzymes is purely catalytic, modifying the rate of assembly but not the final disposition of nucleosomes on DNA. In the other, ATP-dependent chromatin remodeling enzymes actively reposition nucleosomes to control access to DNA, in analogy with motor proteins. The relative importance of intrinsic sequence preferences, chromatin remodeling enzymes, competition with other factors, and formation of higher-order structures for shaping and maintaining *in vivo* chromatin continues to be debated.

To address these questions, large-scale maps of nucleosome positions have been generated in recent years. Using microarray and massively parallel sequencing technologies, nucleosomes have been mapped genome-wide in *S.cerevisiae* (both *in vivo* and *in vitro*), *D.melanogaster*, *C.elegans* and *H.sapiens*. These data were used to train bioinformatics models that attempt to predict nucleosome occupancy profiles and in particular discriminate between nucleosome-enriched and depleted regions using various sequence features (*e.g.* dinucleotide frequencies found in the alignment of experimentally mapped nucleosomal sequences (Segal *et al.* 2006)) as input. On the other hand, there are a number of models that do not rely on high-throughput data sets – rather, they employ a physics-based description of the nucleosome core particle to predict sequence-dependent elastic energies of bending nucleosomal DNA into a superhelix. The elastic energy is represented as a quadratic potential, with empirical parameters obtained from DNA structural data, measurements of DNA mobility on a gel, etc. The resulting energy profile is then used to predict nucleosome occupancies and positions and compare them with experimental data.

In this review we focus on high-throughput nucleosome positioning data sets and on the computational models created to explain them. The scope of this review does not allow us to discuss smaller-scale nucleosome positioning studies that had existed well before the first microarray-based nucleosome map was published in 2005. Although we have tried to review all major data sets published between 2005 and 2009, rapid progress in parallel sequencing technology makes it certain that many more nucleosome positioning maps will be available in the near future. The current data sets however have already been invaluable for understanding major features of nucleosome organization in eukaryotic genomes. They have also provided insight into how chromatin is shaped by genomic sequence features that dictate nucleosome positioning *in vitro*. In the modeling part of the review we focus on bioinformatics models and on physical models that compute sequence-dependent free energies of nucleosome formation.

This review is organized as follows: Section II.1 covers nucleosome positioning data and describes select observations drawn from it, focusing especially on stereotypical features of nucleosome organization in genic and intergenic regions. Section II.2 gives a short overview of the structural studies of the nucleosome core particle. Section III.1 describes nucleosome models based on elastic energy calculations. Section III.2 summarizes currently available bioinformatics models for chromatin structure prediction. Section III.3 explains how statistical physics of one-dimensional (1D) liquids can be applied to positioning multiple nucleosomes simultaneously without steric overlap. Section III.4 describes a Hidden Markov Model (HMM) approach for inferring nucleosome occupancy from log-intensity microarray profiles. Finally, Section IV contains a brief summary of our main observations.

## II. Experimental studies of chromatin structure

## 1. Genome-wide mapping of nucleosome positions

### 1.1. Nucleosome positioning studies in *Saccharomyces cerevisiae*

### 1.1.1. Microarray studies

Given a current wealth of technologies for genome-wide mapping of nucleosome positions, it is amazing to note that these technologies date only from 2005. Prior to that, microarray resolution was simply too low (~1 kbp) to detect single nucleosome positions. Despite this limitation, early work showed general depletion of nucleosomes from promoter regions (Bernstein *et al.* 2004; Lee *et al.* 2004). This important observation was refined in a pioneering 2005 study by Yuan *et al.* (Yuan *et al.* 2005) that employed a microarray to map nucleosome positions across 482 kb of the budding yeast genome, spanning almost the entire chromosome III and 223 additional regulatory regions. The microarray consisted of 50 bp oligonucleotide probes tiled every 20 bp across the genomic regions of interest. Nucleosomal DNA was digested with micrococcal nuclease (MNase; an endo-exonuclease that preferentially digests linker DNA) and labeled with green (Cy3) fluorescent dye. The sample was then mixed with total genomic DNA labeled with red (Cy5) fluorescent dye and hybridized to the microarray. The output of the experiment consisted of the log2 ratio of hybridization values for nucleosomal versus genomic DNA at each probe position across the genomic region of interest (see Figure 2 for the overview of the method).

The log-intensity profiles served as input to a Hidden Markov Model (HMM) (Section III.4) which was used to predict probabilities of starting a nucleosome at each bp and the corresponding nucleosome occupancies (Nucleosome occupancy is defined as the probability that a given bp is covered by any nucleosome and is therefore computed as the sum of probability peaks for all nucleosomes that are close enough to overlap the bp in question). The authors confirmed earlier low-resolution reports that intergenic DNA in yeast was nucleosome-depleted relative to coding DNA, and found nucleosome-depleted regions (NDRs) of ~150 bp in length immediately upstream of many annotated coding sequences (Figure 3). Although microarray resolution was insufficient for mapping individual nucleosomes with bp precision, the authors were able to carry out a limited study of sequence determinants of nucleosome positioning, and found that nucleosome-free regions were enriched in poly-A and poly-T motifs. These motifs tend to occur in promoters, suggesting a causal role of poly(dA-dT) tracts in establishing NDRs.

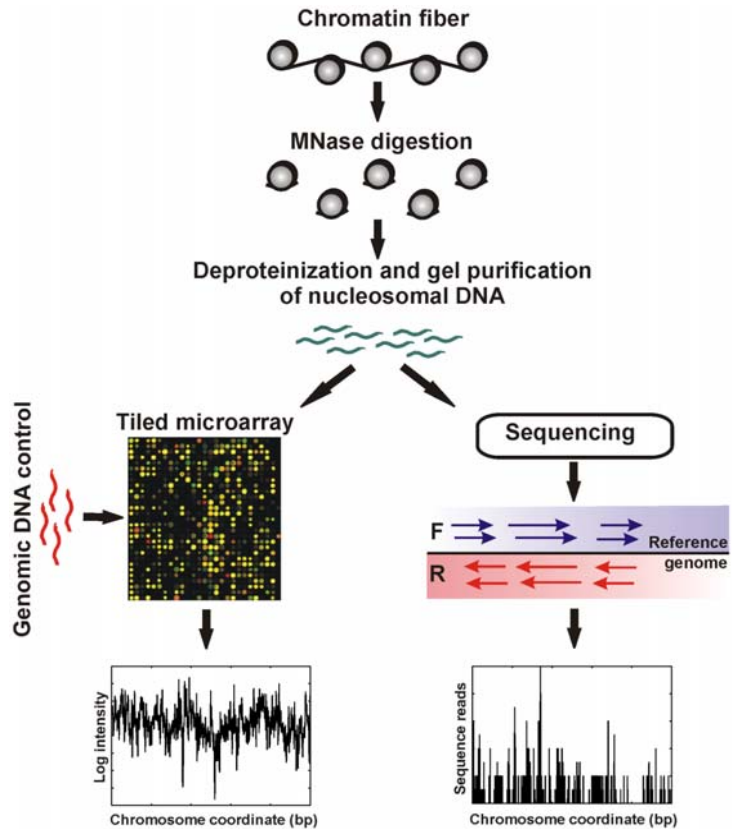

**Figure 2** Schematic representation of high-throughput nucleosome positioning experiments. Chromatin is subjected to MNase digestion, the resulting DNA is purified and mononucleosomal DNA is isolated on a gel. Optionally, histones are cross-linked on DNA prior to MNase digestion and immunoprecipitated, after which the cross-linking is reversed. In a chip-based experiment (left panel), mononucleosomal DNA is hybridized to a microarray. A separate hybridization is carried out for a control sample prepared by digesting genomic nucleosome-free DNA with MNase. The log ratio of nucleosomal DNA intensity to genomic DNA intensity is plotted as a function of each probe's starting position on the genome. Higher than average log-intensity values correspond to nucleosome-covered regions, while lower than average log-intensity values correspond to nucleosome-depleted regions. Microarray probes are tiled across the entire genome or genomic regions of interest. In a high-throughput parallel sequencing experiment, mononucleosomal (and in some cases control) DNA is sequenced directly (right panel). High-throughput sequencing yields large collections of reads that are typically shorter than DNA lengths in the input sample. The reads are mapped onto both strands of the reference genome (often with several mismatches allowed) and combined into a single sequence read profile by assuming that each nucleosome core particle has a fixed length of 147 bp. With this assumption, sequence reads from the reverse strand are remapped onto the forward strand by subtracting 147 bp from their end coordinates. The resulting sequence read profile contains information about the number of reads assigned to every genomic position.

Rapid progress in microarray technology allowed the first nucleosome map of the entire *S.cerevisiae* genome to be completed in 2007 (Lee *et al.* 2007). The Lee *et al.* study employed high-density Affymetrix tiling microarrays with 25 bp probes spaced every 4 bp across the yeast genome. Similarly to the earlier Yuan *et al.* study (Yuan *et al.* 2005), genomic chromatin was cross-linked with formaldehyde and treated with MNase, resulting in preferential digestion of nucleosome-free linker sequences. Three independent samples of nucleosomal and total genomic DNA were hybridized to the microarray. The resulting log2-intensity traces were processed by HMM to yield genome-wide nucleosome probabilities and occupancies. The authors used the Viterbi algorithm (Durbin *et al.* 1998) to identify nucleosome positions in the "top" (most likely) nucleosome configuration. Using a single configuration rather than a sum over all possible configurations allows placing nucleosomes uniquely but discards probabilistic information about alternative nucleosome positions contained in the full HMM approach (Section III.4). The authors confirmed existence of NDRs on the genome-wide scale and observed that nucleosome occupancy profiles correlated with transcript abundance and transcription rate. In addition, functionally related genes could be clustered on the basis of nucleosome occupancy patterns observed at their promoters.

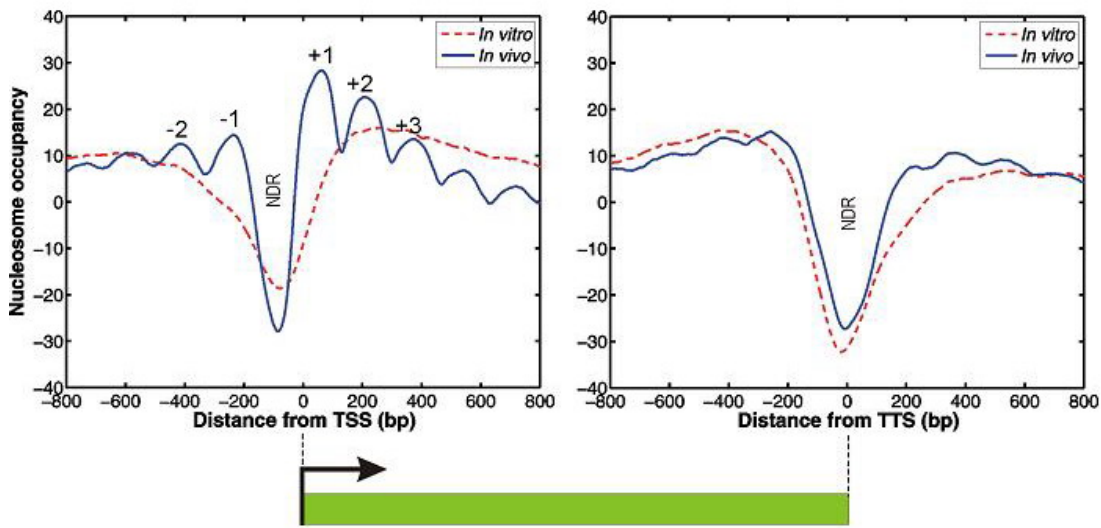

**Figure 3** Nucleosome occupancy in the vicinity of transcription start and termination sites in *S.cerevisiae*. The unnormalized occupancy is defined as the number of nucleosomes covering a given bp. Blue solid lines: *in vivo* occupancy (YPD medium, average over four replicates without cross-linking), red dashed lines – *in vitro* occupancy (average over two replicates). *In vivo* and *in vitro* nucleosome positions were mapped by Kaplan *et al.* using high-throughput sequencing (Kaplan *et al.* 2009). Transcript coordinates are from Nagalakshmi *et al.* (Nagalakshmi *et al.* 2008). The genome-wide average of the nucleosome occupancy is subtracted from the plots.

## 1.1.2. High-throughput sequencing studies

By 2007 massively parallel sequencing technologies had matured to the point where it became possible to sequence hundreds of thousands of nucleosomal DNA molecules directly instead of hybridizing them to a microarray (see Figure 2 for an overview of the method). The necessary technology, first developed by 454 Life Sciences (http://www.454.com), was capable of sequencing ~100 bp molecules, comparable to the 147 bp length of the nucleosome core particle. Thus it was natural to apply high-throughput sequencing to nucleosomal DNA. The first such study, carried out by Frank Pugh and co-workers, focused on *S.cerevisiae* nucleosomes with the histone variant H2A.Z (Albert *et al.* 2007). H2A.Z is conserved by evolution and is believed to be involved in transcriptional regulation, antisilencing, silencing, and genome stability (Draker and Cheung 2009).

The Albert *et al.* study mapped 322,000 H2A.Z nucleosomes that had been trapped by formaldehyde cross-linking, immunoprecipitated, MNase-digested, and gel-purified. The most important difference with the microarray experiments was that instead of hybridizing nucleosomal DNA samples to the array (after further fragmenting the ~150 bp sequences with nuclease into ~50 bp pieces to make hybridization with relatively short probes more efficient), DNA molecules were sequenced directly and mapped onto the reference genome, providing a starting coordinate for each sequence read. This procedure resulted in a sequence read profile in which a non-negative number of reads was assigned to each genomic bp (Figure 2). Unlike microarray hybridization approaches, sequence read profiles can be constructed separately for both strands and either compared with each other (Figure 4) or combined. The authors had to assume that the nucleosome position was given precisely by the coordinates of the sequence read, neglecting potential errors caused by imperfect MNase digestion. They also had to extend mapped sequence reads to cover the 147 bp length of the nucleosome core particle (this procedure is common to all short-read sequencing studies). Furthermore, because only H2A.Z – containing nucleosomes were detected, the question of bulk nucleosome positioning remained open.

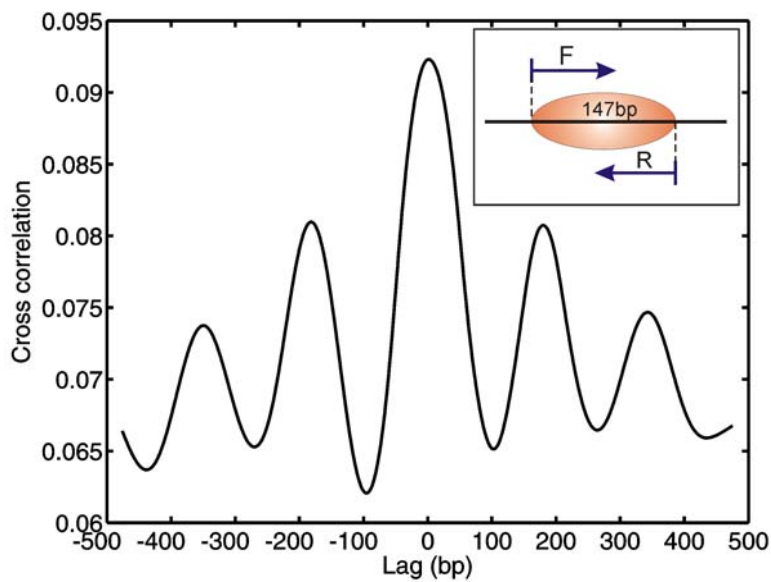

**Figure 4** Correlation function between sequence reads mapped onto Watson and Crick strands. High-throughput sequencing data is for *in vivo* nucleosomes in *S.cerevisiae* (Mavrich *et al.* 2008a). The observed maximum at zero lag corresponds to sequence reads on the opposite strands separated by 147 bp and thus demarcating the same nucleosome (see inset). Maxima at non-zero lag show relative positions of neighboring nucleosomes. The correlation function was smoothed using a 50-bp window average.

This question was addressed in a 2008 study from the same group (Mavrich *et al.* 2008a) which used immunoprecipitation with antibodies against tagged histones H3 and H4 to map 1,206,057 bulk nucleosomes from the yeast genome. The use of 454 sequencing and the construction of the sequence read profile were the same as in the earlier H2A.Z study. These sequencing studies helped establish a "canonical" picture of nucleosome organization in which well-positioned -1 and +1 nucleosomes bracket an NDR upstream of *S.cerevisiae* genes (Figure 3).

Mavrich *et al.* argued that positioning of bulk nucleosomes is largely a consequence of steric exclusion: +1 and to a certain extent -1 nucleosomes form a barrier against which the other nucleosomes are "phased". In this scenario, sequence specificity would be important for only a small fraction of positioned nucleosomes. This picture is consistent with the observation that nucleosomal dinucleotide patterns are more pronounced in the -1 and +1 nucleosomes than in the bulk ones (Mavrich *et al.* 2008a). The observed patterns are non-periodic and consist of a gradient of the TA/AA/TT dinucleotide counts at positions covered by -1 and +1 nucleosomes. The gradient is absent from bulk nucleosomes and its direction corresponds to a decrease in the TA/AA/TT counts with distance to the NDR (that is, the counts increase in the 3' direction for the -1 nucleosome and in the 5' direction for the +1 nucleosome).

In addition to the 5' NDR, the authors discovered a novel 3' NDR (Figure 3) which coincides with the transcription termination site (TTS) and argued that it may be implicated in transcription termination, anti-sense initiation, and gene looping. They concluded that the terminal nucleosomes (3' nucleosomes immediately upstream of the TTS) may be partially positioned by sequence, including nearby cleavage and polyadenylation sites (AATAAA and related sequences).

Another *in vivo* map of nucleosome positions in *S.cerevisiae* was published by Eran Segal and co-workers in 2008 (Field *et al.* 2008). The authors used 454 parallel pyrosequencing technology to sequence 503,264 yeast nucleosomes. The nucleosomes were mapped to the yeast genome by BLAST with the 95% sequence identity cutoff. The authors also required that sequence reads map to a unique location, have a length between 127 and 177 bp, and do not overlap with the ribosomal RNA locus. The resulting 378,686 nucleosomes were retained for further analysis. 454 pyrosequencing technology used in this study was capable of creating ~200 bp reads (longer than the ~100 bp reads obtained by Frank Pugh and colleagues (Albert *et al.* 2007; Mavrich *et al.* 2008a)) and thus mononucleosome fragments were sequenced in full. However, although both ends of the mononucleosome read were known, a wide distribution of fragment lengths made it impossible to predict individual nucleosome positions with bp precision. Furthermore, there was no immunoprecipitation step, and the sequence read coverage was approximately a quarter of the 454 data set created by Frank Pugh and colleagues (Mavrich *et al.* 2008a).

The main focus of the Field *et al.* paper is on the nucleosome positioning signals, which the authors captured with a model based on: a) relative dinucleotide frequencies at 127 central positions in the alignment of all sequence reads around their center (Segal *et al.* 2006); b) ratios of 5-mer frequencies in the linkers (defined as contiguous regions of 50-500 bp not covered by any nucleosome) and in the nucleosome-bound sequences. The model is described in more detail in Section III.2.1; here we simply note that AAAAA/TTTTT was found to be the 5-mer with the strongest enrichment in linkers. The authors attribute a significant nucleosome positioning role to this and other A/T-rich "boundary zone" elements that tend to be under-represented in nucleosomes but over-represented immediately outside of the nucleosome cores in the linker regions.

The authors also argue that nucleosome depletion over A/T rich "boundary elements" is unlikely to be an MNase artifact, on the basis of the low rank-order correlation observed between word frequencies across MNase cut sites and relative word frequencies in linkers vs. nucleosomes. In particular, AAAAAA is ranked 1782th as an MNase cleavage site and 1st for enrichment in linker regions. The authors also find that their data yields an average nucleosome occupancy profile with respect to the transcription start site (TSS) that is in a broad agreement with the earlier microarray study by Lee *et al.* (Lee *et al.* 2007): for most genes, there is a prominent NDR flanked by oscillations in the nucleosome occupancy that are usually interpreted as a consequence of steric exclusion.

Except for several studies focused on changes in chromatin structure with respect to environmental or genetic perturbations (cf. Section II.1.1.3), prior to 2009 all nucleosome maps had come from *in vivo* chromatin of yeast cells grown in rich YPD medium. However, in 2009 Eran Segal and co-workers employed short-read Solexa/Illumina sequencing (http://www.illumina.com/sequencing) to compare *in vitro* and *in vivo* nucleosome positions and to study how chromatin structure changes under different growth conditions (Kaplan *et al.* 2009). The *in vitro* map is especially important because nucleosome locations should be dictated purely by steric exclusion and intrinsic sequence preferences. For *in vivo* maps, yeast cells were grown in YPD medium as well as YP media supplemented with 2.0% galactose or 2.8% ethanol instead of glucose. For each medium, nucleosome DNA samples were prepared both with and without formaldehyde cross-linking and sequenced-by-synthesis using the Solexa/Illumina technique. For the *in vitro* map, yeast genomic DNA was purified and mixed with histone octamers from chicken erythrocytes. Nucleosomes were reconstituted by salt gradient dialysis (Thastrom *et al.* 2004) at the lower histone octamer concentration than that observed *in vivo* (40 ìg histone octamer per 100 ìg DNA). The lower histone concentration was necessary as reconstitutions at higher *in vivo* stoichiometry resulted in insoluble chromatin which was inaccessible to MNase.

Comparison between nucleosome positions from *in vitro* and *in vivo* experiments revealed striking overall similarity, leading the authors to conclude that: a) *in vitro* and *in vivo* nucleosome maps are highly similar; b) chromatin structure is largely invariant with respect to different growth conditions. The authors also concluded that nucleosome positions are largely encoded by intrinsic DNA sequence signals, because a purely sequence-dependent model fit on the *in vitro* data was able to predict *in vivo* nucleosome locations with high accuracy. The model is essentially identical to the earlier one from Field *et al.* (Field *et al.* 2008), and is described in more detail in Section III.2.1. Similarly to this previous study, the authors found using both *in vitro* and *in vivo* maps that 5-mers with the lowest average nucleosome occupancy were AAAAA and ATATA. In addition to the nucleosome-excluding and nucleosome-favoring distributions of 5-mers, the authors described a 10-11 bp periodic dinucleotide signal caused by DNA bending (as discussed in the Introduction), with AA/AT/TA/TT frequencies out of phase with CC/CG/GC/GG frequencies (Figure 5). These periodic oscillations have been observed in both *in vitro* and *in vivo* nucleosome positioning sequences (Albert *et al.* 2007; Field *et al.* 2008; Mavrich *et al.* 2008a; Kaplan *et al.* 2009).

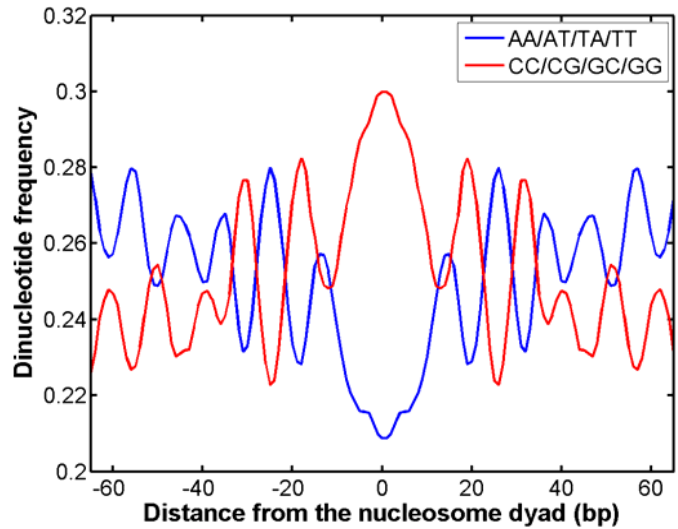

**Figure 5** Periodic dinucleotide frequencies observed in a high-throughput data set of nucleosome positioning sequences in yeast (Kaplan *et al.* 2009). 147 bp long *in vitro* nucleosome positioning sequences defined by five or more sequence reads were aligned and the relative frequencies of AA/AT/TA/TT and CC/CG/GC/GG dinucleotides were computed at each position in the nucleosomal site. The frequencies were divided by genome-wide propensities for each group of dinucleotides.

Finally, the nucleosome-depleted region is present *in vitro* at both the TSS and the translation end (which was chosen because the authors believe transcription termination sites to be poorly annotated in yeast). Interestingly, while there is no difference between *in vivo* and *in vitro* nucleosome depletion at 3' ends of genes, the 5' NDR is much shallower for *in vitro* chromatin and there are no characteristic oscillations in the flanking regions (Figure 3). If these oscillations are indeed induced by steric exclusion, their absence from *in vitro* chromatin indicates that nucleosomes are not positioned as precisely, and suggests that intrinsic sequence signals are not the only contribution to *in vivo* anchoring of nucleosomal arrays.

Another study which focused on mapping nucleosomes assembled *in vitro* on genomic DNA was carried out by Kevin Struhl and co-workers (Zhang *et al.* 2009). The authors purified both *S.cerevisiae* and *E.coli* genomic DNA and assembled it into chromatin either by salt dialysis with *D.melanogaster* histones or by using an *in vitro* system containing recombinant *D.melanogaster* proteins NAP-1 (nucleosome assembly protein 1) and ACF, an ATP-dependent chromatin assembly factor known to produce arrays of regularly spaced nucleosomes (Ito *et al.* 1997). *In vitro* chromatin was subsequently digested with MNase to mononucleosome core particles, the resulting DNA was purified and sequenced on a Solexa/Illumina Genome Analyzer, yielding 1 to 3 million uniquely mapped sequence reads for each input sample. As a control, the authors sonicated the mixture of yeast and *E.coli* DNA to fragments of mononucleosomal size.

The authors discovered that nucleosomes strongly prefer yeast DNA to *E.coli* DNA, indicating that yeast genome evolved to facilitate nucleosome formation. As in the Kaplan *et al.* study (Kaplan *et al.* 2009), they found that many regions around transcription start and termination sites intrinsically disfavor nucleosome formation, and that nucleosomes positioned *in vitro* by salt dialysis exhibit prominent periodic distributions of AA/TT/AT dinucleotides. In contrast, nucleosomes positioned by ACF had fewer NDRs and significantly less prominent periodicity of dinucleotide frequencies, showing that ACF is capable of overriding intrinsic sequence specificities of nucleosome core particles.

The main conclusion reached by Zhang *et al.* (Zhang *et al.* 2009) is that intrinsic histone-DNA interactions are not a major determinant of *in vivo* nucleosome positions. This is in contrast with the Kaplan *et al.* study (Kaplan *et al.* 2009) which argues that *in vivo* and *in vitro* nucleosome occupancy profiles are highly similar and that the latter can be explained with a model based purely on DNA sequence features. Zhang *et al.* focus instead on the fact that the *in vivo* pattern of statistical nucleosome positioning around 5' NDRs is not observed *in vitro* and thus cannot be determined by intrinsic sequence preferences alone (Figure 3). They argue that strong positioning of the +1 nucleosome is linked to the process of transcriptional initiation, and propose that, although *in vitro* 5' NDRs may facilitate assembly of the pre-initiation complex, an early step in the transcription process (probably preceding extensive elongation) is a primary determinant of the +1 nucleosome positioning. They hypothesize that some component of the transcriptional initiation machinery interacts with a nucleosome-remodeling complex and/or histones to position the +1 nucleosome. Once in place, the +1 nucleosome positions +2,+3,… nucleosomes by steric exclusion. This view is supported by the observation that *in vivo* +2,+3,… nucleosomes are much better positioned than their -2,-3,… counterparts, even though the intrinsic positioning effect of the NDR should be the same on both sides.

### 1.1.3. Physiological and genetic perturbations of *S.cerevisiae* chromatin

There are several studies of nucleosome positions in yeast whose main focus is not on intrinsic sequence preferences or on the nucleosome organization with respect to various genomic features, but rather on how chromatin responds either to environmental perturbations or to deleting genes implicated in chromatin remodeling and maintenance.

The first high-throughput study of this kind was carried out in 2007 by Toshio Tsukiyama and co-workers (Whitehouse *et al.* 2007). The authors investigated the role of the ATP-dependent chromatin remodeling complex Isw2 in controlling chromatin structure across the yeast genome. They sought to discover Isw2 targets genome-wide by identifying differences in nucleosome positions between wild-type and $\Delta isw2$ mutant strains. To this end, chromatin was cross-linked by formaldehyde, digested with MNase and exonuclease III, and purified to mononucleosomes using gel electrophoresis. Mononucleosomal DNA from both strains was separately hybridized to high-resolution Affymetrix tiling microarrays with ~5 bp probe spacing. Based on the difference in hybridization intensity between wild-type and mutant strains, the authors identified >1,000 regions, typically ~600 bp in length, where chromatin structure was disrupted in the $\Delta isw2$ mutant (these changes affected ~12% of yeast promoters). The authors also used chromatin immunoprecipitation (ChIP) to determine whether Isw2 was present at the loci whose nucleosome positions changed between strains.

The authors concluded that Isw2 functions by moving nucleosomes towards intergenic regions, where many important regulatory sequences are located. By doing so, it overrides intrinsic sequence-based nucleosome positioning signals, as evidenced by the fact that poly(dA-dT) tracts are located within nucleosome +1 at many Isw2 targets (defined as promoters that exhibited change in the $\Delta isw2$ deletion strain). Loss of Isw2 would thus allow nucleosomes to relocate to their inherently preferred sites, lowering the total free energy of the system. The ability of Isw2 and other chromatin remodeling enzymes to actively reposition nucleosomes demonstrates that intrinsic nucleosome positioning preferences may be disrupted in living cells.

The Whitehouse *et al.* Isw2 study (Whitehouse *et al.* 2007) was followed in 2009 by Hartley and Madhani (Hartley and Madhani 2009) who focused on how nucleosome positioning was affected by degrading proteins believed to be essential for maintaining chromatin structure: Myb family proteins Abf1 and Reb1 and the catalytic subunit of the RSC remodeling complex, Sth1. Since Reb1, Abf1 and Sth1 are all essential proteins, the authors had to use conditional alleles rather than gene deletion strains. Specifically, they used the temperature-sensitive degron system to engineer yeast strains in which protein degradation could be controlled via the N-end rule pathway (Dohmen and Varshavsky 2005). Nucleosomal DNA samples were collected from wild-type and "degron" strains and interrogated using a low-resolution microarray designed to cover yeast chromosome III at 20 bp tiling steps (Yuan *et al.* 2005). The authors found that while the effect of Reb1 and Abf1 depletion was minimal, affecting ~10% of chromosome III promoters, depleting Sth1 affected the majority (~55%) of promoters. The affected genes displayed shrinking of the NDR accompanied by the movement of flanking nucleosomes. Although NDR was reduced in width, it was not eliminated, and the authors hypothesized that it is maintained by intrinsic sequence preferences. In support of this hypothesis, nucleosome positions were better predicted by the intrinsic "nucleosome positioning signature" (a first-generation nucleosome positioning model by Ioshikes *et al.* (Ioshikhes *et al.* 2006), cf. Section III.2.2) in the absence of Sth1.

It is also of interest to know how chromatin structure in yeast cells responds to physiological perturbations such as heat shock that are usually accompanied by massive transcriptional changes. The first study of this kind was carried out in 2008 by Vishwanath R. Iyer and colleagues (Shivaswamy *et al.* 2008). The authors subjected yeast cells grown in rich medium to a 15-min period of heat shock. At the end of the 15-min period, control and heat-shocked cells were treated with formaldehyde. Mononucleosomal DNA was isolated by means of a standard protocol which involves MNase digestions and gel purification, and sequenced using Solexa/Illumina short-read technology. In this way, the authors generated a differential map of nucleosome positions which consisted of 514,803 and 1,036,704 uniquely mapped reads for the normal and heat-shock growth conditions, respectively.

As in work by Frank Pugh and colleagues which was published at approximately the same time (Mavrich *et al.* 2008a), the authors find both 5' and 3' NDRs, with a well-positioned nucleosome at the 3' end of the coding region. Thus, yeast genes are demarcated by NDRs at each end of the transcribed region. Nucleosomes located next to the NDRs are well-positioned, at least in part because NDRs act as barriers against which the genic nucleosomes are "phased".

The authors also addressed a question of whether positioning of terminal or bulk nucleosomes over transcribed regions could be attributed to intrinsic sequence signals. They discovered that the distribution of AA/TT dinucleotide frequencies is 10-11 bp periodic in +1 nucleosomes. Furthermore, this periodic profile is also observed in +2, +3, … nucleosomes, showing that their positioning may be maintained through sequence signals in addition to steric constraints.

Surprisingly, the majority of nucleosomes did not change positions upon transcriptional perturbation caused by heat shock, either in promoters or in the coding regions. At some promoters, remodeling events were observed that could be classified into eviction, appearance, or repositioning of one or two nucleosomes. However, there were no simple rules that controlled nucleosome remodeling at induced and repressed promoters. Thus, although gene activation was associated on average with nucleosome eviction and gene repression with nucleosome appearance, there were cases in which strongly positioned nucleosomes appeared at induced promoters. Furthermore, many nucleosome remodeling events occurred at promoters that did not experience a significant transcriptional change.

Many of these findings were corroborated in a 2009 study in which global nucleosome positioning was examined before and after global transcriptional restructuring caused by adding glucose to yeast cells grown in a poor carbon source (Zawadzki *et al.* 2009). This nutrient upshift creates significant changes in gene expression of more than half of all yeast genes (Zaman *et al.* 2009). The authors isolated mononucleosomal DNA using standard MNase digestion methods at three time points: immediately before, 20 min, and 60 min after adding glucose to yeast cells. Nucleosomal and total genomic DNA were hybridized to Affymetrix microarrays tiled each 4 bp across the yeast genome. The authors also developed a Hidden Markov Model approach to processing log-intensity profiles (cf. Section III.4) which they used to predict nucleosome occupancies across the yeast genome.

The authors discovered that for most genes changes in expression were not associated with nucleosome addition, removal, or repositioning within their promoters, although for genes containing TATA boxes the correlation between change in gene expression and change in nucleosome occupancy was somewhat higher (0.48 for TATA genes vs. 0.34 for all genes). The promoters of only 10% of all genes gained or lost nucleosomes despite the fact that >50% of all genes exhibited a change in mRNA levels of twofold or more (Zaman *et al.* 2009). Thus it appears that *in vivo* interactions of transcription factor (TF) binding sites with their cognate factors are largely dictated by prepositioned nucleosomes and that regulation of gene expression through these sites is mediated by changes in local TF concentration rather than nucleosome addition or removal. The unaltered promoter nucleosome structure for most glucose-regulated genes implies the existence of constitutively accessible binding sites for the factors that control expression of these genes. This is consistent with the notion of "preset" chromatin which plays a largely instructive role in regulating gene expression (Morse 2007).

## 1.2. Nucleosome positioning studies in higher eukaryotes

### 1.2.1. *Caenorhabditis elegans*

High-throughput sequencing and microarray methods pioneered with *S.cerevisiae* were soon applied to other eukaryotes. In two recent papers, Andrew Z. Fire and co-workers mapped nucleosome positions in another model organism, *C.elegans* (Johnson *et al.* 2006; Valouev *et al.* 2008). The first of these studies presented a collection of 284,091 nucleosome cores sequenced with the 454 pyrosequencing technology. The nucleosomes came from a mixed-stage population of *C.elegans*. As in yeast, nucleosome sequence reads were mapped onto the reference genome, with ~60% of all reads assigned unambiguously to genomic loci. The resulting map had the coverage of one nucleosome per 300-400 bp of genomic DNA. Analysis of dinucleotide distributions revealed a pronounced periodicity in AA and TT frequencies which extended across the nucleosome core.

Another observation concerned the effect of the MNase sequence specificity on nucleosome positioning: in agreement with the earlier study of MNase sequence preferences (Wingert and Von Hippel 1968), the authors observed preferential cleavage at A/T-rich target sites, with G/C residues considerably underrepresented at both positions flanking the cleavage site. However, there is no corresponding A/T enrichment around position 147 (where the end of the nucleosome core particle would be if it were mapped with bp precision). The authors interpret this lack of symmetry as evidence that MNase sequence specificity influences the choice of cleavage sites but does not lead to nucleosome repositioning. Indeed, if nucleosomes were actively repositioned by interactions with MNase, both ends of the core particle would be marked by A/T-rich flanking regions. As mentioned above, symmetric flanking regions were not observed in the data, although this may simply be due to inaccuracies in locating the other end of the nucleosome.

This study was extended in 2008 by the same group using a massively parallel technique of sequencing by oligonucleotide ligation and detection (SOLiD by Applied Biosystems: solid.appliedbiosystems.com) (Valouev *et al.* 2008). Parallel sequencing yielded more than 44 million uniquely mapped nucleosome cores from a mixed-stage population of *C.elegans*. SOLiD sequencing platform produces 50 bp reads - shorter than the 147 bp length of the nucleosome core particle. As a result (and similarly to other sequencing studies), the position of the nucleosome dyad had to be inferred by adding 73 bp to the starting bp for reads mapped onto the Watson strand and subtracting 73 bp from the starting bp for reads mapped onto the Crick strand.

The authors found that absolute nucleosome positions varied substantially, possibly reflecting a lack of universal sequence-dictated positioning across *C.elegans* cell types. Nonetheless, nucleosomes tended to be arranged in repeated array-like structures, presumably due to steric constraints. Sequence analysis of nucleosome cores showed an oscillating ~10 bp periodicity for AA/TT with an out-of-phase ~10 bp periodicity for GC. Longer words (up to 6-mers) were distributed non-randomly as well, with a pronounced enrichment of A/T nucleotides around sequence read starts which the authors again ascribed to MNase sequence specificity.

### 1.2.2. *Drosophila melanogaster*

In 2008 Frank Pugh and co-authors used 454 pyrosequencing to map 652,738 H2A.Z-containing nucleosomes to 207,025 locations in the *D.melanogaster* genome (Mavrich *et al.* 2008b). Similarly to the yeast studies from the same group (Albert *et al.* 2007; Mavrich *et al.* 2008a), *Drosophila* embryos were treated with formaldehyde, H2A.Z-containing nucleosome core particles were immunopurified and nucleosomal DNA was sequenced. Because *Drosophila* embryos consist of a wide variety of cell types, the nucleosome map is an average over cells with potentially very different gene expression profiles. Nonetheless, the nucleosome organization showed generic features that transcended the differences in cell types. The most prominent of these was the nucleosome-depleted region upstream of the TSS. There were two essential differences with the earlier study of H2A.Z nucleosomes in yeast (Albert *et al.* 2007): the absence of a well-positioned -1 nucleosome and longer linker lengths in the fly, manifested by larger distances between consecutive nucleosomal peaks downstream of the +1 nucleosome. In addition, the genic (+1, +2, ...) array of nucleosomal peaks started ~75 bp further downstream from the equivalent position in *S.cerevisiae*, with potentially important implications in how the TSS is presented to RNA polymerase II (Pol II): in *S.cerevisiae* the TSS resides within the nucleosome border, whereas in *D.melanogaster* the TSS tends to be nucleosome-free. This is consistent with the hypothesis that gene regulation occurs predominantly at the level of transcript initiation in *S.cerevisiae*, whereas in *D.melanogaster* transcript elongation may play a more important role.

As in yeast, the 3' ends of fly genes tend to be nucleosome-depleted. H2A.Z nucleosome positioning sequences exhibit periodic, out-of-phase distributions of A/T-rich and G/C-rich dinucleotides. Finally, there is a correlation between AA/TT and CC/GG content and nucleosome positioning: nucleosome-covered positions tend to be G/C-rich, whereas 5' and 3' NDRs are enriched to some extent in A/T nucleotides, including poly(dA-dT) transcription termination sites at the 3' end of *Drosophila* genes. The same study produced a lower-resolution map of bulk (both H2A and H2A.Z-containing) nucleosomes by digesting chromatin with MNase and hybridizing DNA samples to Affymetrix *Drosophila* tiling arrays with 36 bp average probe spacing (there was no immunoprecipitation step in this assay). The same nucleosome positioning pattern was found with respect to the TSS, with the exception of a distinct -1 peak which was not present in the H2A.Z map. Thus in both yeast and fly -1 nucleosomes are well-positioned but in the fly they tend to be H2A.Z-free, resulting in the absence of a prominent -1 peak in H2A.Z maps.

### 1.2.3. *Homo sapiens*

The first study of nucleosome positioning on human genome was carried out in 2007 using high-resolution microarrays. Using MNase digestion, Ozsolak *et al.* isolated mononucleosomal DNA from five types of human cells: primary fibroblasts (IMR90), primary melanocytes (PM), mammary epithelial cells (MEC), melanoma (A375, MALME), and breast cancer cell lines (T47D, MCF7) (Ozsolak *et al.* 2007). Nucleosome-free genomic DNA from the same cell line (digested to a similar size distribution) was used as control. Nucleosomal and genomic DNA samples were labeled by different fluorescent dyes (Cy5 and Cy3, respectively) and hybridized to microarrays containing 50 bp probes. The probes were tiled in 10 bp steps and spanned 1.5 kb repeat-masked promoter regions of 3,692 genes, including 1,346 genes in the Affymetrix Human Cancer G110 Array and 2,346 randomly selected genes.

Because the log-intensity profiles came in 1.5 kb fragments, the HMM algorithm could not be used for predicting nucleosome positions and occupancies (HMMs require contiguous input data (Durbin *et al.* 1998)). The authors chose to use wavelet-based de-noising instead, followed by an edge-detection algorithm. Because log-intensity microarray profiles measure nucleosome occupancies, application of edge-detection techniques amounts to predicting nucleosome positions from occupancy data.

In order to examine whether NDRs existed in human cells, Ozsolak *et al.* compared nucleosome organization in expressed and unexpressed promoters from A375, IMR90 and MALME cell lines. On average, NDRs of expressed genes were much more pronounced. Some of the unexpressed genes also had NDRs, which the authors attributed to the fact that chromatin structure was pre-modified, making those genes poised for rapid expression. This hypothesis was supported by an observation that NDRs of unexpressed genes were likely to have transcription pre-initiation complexes (PICs) pre-assembled at their promoters. Conversely, unexpressed genes without PICs had no nucleosome depletion around the TSS.

Finally, the authors looked for short sequence motifs preferentially enriched or depleted in nucleosome-covered regions. They found that TATAAA, TATATA, GCGCGC and AAAAAA motifs were enriched in nucleosome linkers, while TTCGA and CTGCTG motifs were enriched in nucleosome cores. The authors argued that since none of the linker-enriched motifs corresponded to the previously published MNase recognition sequences (Horz and Altenburger 1981), MNase sequence specificity did not exert a significant influence on detected nucleosome positions, nor did it bias which subset of nucleosomes is detected. This conclusion is similar to that reached in yeast by Field *et al.* (Field *et al.* 2008), but is at variance with the *C.elegans* studies (Johnson *et al.* 2006; Valouev *et al.* 2008).

Another nucleosome positioning experiment was published later in 2007 by Robert E. Kingston and co-workers (Dennis *et al.* 2007). The authors employed two complementary approaches for mapping nucleosomes in human genome: a tiling microarray and a capillary electrophoresis-based sequencing. The microarray was custom-designed by NimbleGen Systems, Inc. (http://www.nimblegen.com) and consisted of 50 bp probes tiled every 20 bp on both forward and reverse strands, with no repeat masking. Three replicates for each strand were spotted on the array. Mononucleosomal DNA and genomic DNA were labeled with Cy3 and Cy5, respectively, and hybridized to the array by the manufacturer. The authors estimate that a single microarray chip constructed in this way can interrogate up to 200 kb of DNA sequence. Capillary electrophoresis-based sequencing can only cover ~10 kb at a time (Dennis *et al.* 2007), making it less suitable for a genome-wide application. Rather, it is a useful method for verifying nucleosome positions obtained by other, potentially less accurate means. The authors carried out a proof-of-principle experiment by designing microarray probes that spanned two relatively short genomic regions: MMTV-LTR and *LCMT2*. The study focused on the consistency between results obtained by the two techniques, and on the good correspondence with previously published nucleosome positions in the MMTV-LTR locus (Richardfoy and Hager 1987; Fragoso *et al.* 1995).

The first high-throughput sequencing study of *H.sapiens* nucleosomes was carried out in 2007 by Keji Zhao and co-workers (Barski *et al.* 2007). The authors employed a Solexa 1G Genome Analyzer to directly sequence ChIP DNA from mononucleosomes generated by MNase digestion of native chromatin. They collected 36 bp long sequence reads for the genome-wide distribution of 20 histone lysine and arginine methylations as well as histone variant H2A.Z. The latter data set with 7.4 million sequence reads is an extension to the human genome of the yeast and fly H2A.Z maps (Albert *et al.* 2007; Mavrich *et al.* 2008b). Similarly to yeast, H2A.Z was found to be enriched in promoter regions both upstream and downstream of the TSS. Furthermore, H2A.Z binding correlated with gene expression, with H2A.Z-containing nucleosomes present at higher levels in promoters of active vs. silent genes.

In the follow-up work published by the same group in 2008, the authors chose resting and activated human CD4+ T cells as a model system (Schones *et al.* 2008). As in previous studies, mononucleosome-sized DNA was isolated from MNase-digested chromatin and sequenced using the Solexa/Illumina short-read technique. 25 bp reads corresponding to the ends of ~150 bp mononucleosome cores were mapped onto the reference human genome. This procedure yielded 154,582,677 uniquely mapped reads in resting cells and 141,931,997 uniquely mapped reads in activated cells. Only those reads were retained for further analysis. The authors found a familiar pattern of phased nucleosomes around the TSS – similarly to yeast and fly, the +1 and -1 nucleosomes (-1 nucleosome is labeled -2 in (Schones *et al.* 2008)) flank an NDR and serve as termini of nucleosome arrays that become progressively less phased with the distance to the TSS.

Nucleosome phasing with respect to TSS was found to be more pronounced in expressed rather than unexpressed genes, consistent with the earlier observation in yeast showing the absence of nucleosome phasing or depletion in a cluster enriched for stress response genes (Lee *et al.* 2007) (stress response genes are not expressed in the YPD medium which was used to prepare the cell culture in the Lee *et al.* study). The authors observed that the position of the +1 nucleosome depends on gene expression: its 5' end peaked at +40 bp with respect to the TSS in active promoters, but only at +10 bp in inactive promoters. This may be due to Pol II binding in the promoter regions of active genes: Pol II peak was found to be located around +10 bp (Barski *et al.* 2007), overlapping with the nucleosome peak in inactive promoters. Finally, the authors found that gene activation by T cell receptor signaling was accompanied to some extent by nucleosome reorganization in promoters and enhancers: there is a consistent difference in the nucleosome occupancies of resting and activated cells for induced and repressed genes.

## 2. Structural studies of the nucleosome core particle

As of 2009, there are approximately 25 nucleosome structures with resolution $\leq 3.0$ Å available in the Protein Databank (PDB; www.rcsb.org). Although the histones in these structures are derived from chicken, mouse, human, yeast and frog, most of them contain the same 146-bp sequence of á-satellite DNA which adopts nearly identical geometries. Several exceptions include a 145 bp sequence (PDB code 2nzd), sequences that are two mutations away from the á-satellite sequences (2cv5, 1kx3, 1aoi), and a 147 bp highest-resolution (1.9 Å) entry with a single bp insertion (1kx5) (Richmond and Davey 2003). In addition, 2fj7 (solved at 3.2 Å resolution) contains a 16 bp poly(dA-dT) tract in its 147 bp long DNA sequence (Bao *et al.* 2006).

There are also several structures with additions and variations, including three nucleosomes derived from 1aoi and complexed with site-specific minor groove-binding ligands, the pyrrole-imidazole polyamides (Suto *et al.* 2003) (1m18, 1m19, 1m1a); crystal structures with histone Sin mutants (Muthurajan *et al.* 2004) (the 1p3 series), histone variants H2A.Z (Suto *et al.* 2000) (1f66) and macroH2A (Chakravarthy *et al.* 2005) (1u35); and a structure with a pyrroleimidazole hairpin polyamide that spans the nucleosomal "supergroove" (Edayathumangalam *et al.* 2004) (1s32).

As can be seen from the above catalog, the vast space of nucleosomal DNA conformations and sequences has not yet been sufficiently sampled by structural studies. Thus understanding the rules linking DNA geometries with nucleosome positioning and free energies of nucleosome formation requires computational modeling guided by available structural data. In particular, the 1.9 Å crystal structure of the nucleosome core particle (Richmond and Davey 2003) reveals the details of a single DNA conformation with unprecedented accuracy. The DNA structure is remarkably different from that observed in non-histone protein-DNA complexes: DNA trajectory has more than twice the curvature of the ideal superhelix with the same radius (41.9 Å) and pitch (25.9 Å) (Figure 6a). Thus DNA is strongly bent, kinked and twisted on average, with DNA segments bent into the minor groove either kinked or alternatively shifted. This is not surprising because DNA is stiff at the length scales of a single core particle and therefore bending it into a superhelix requires major conformational distortions of free B-DNA.

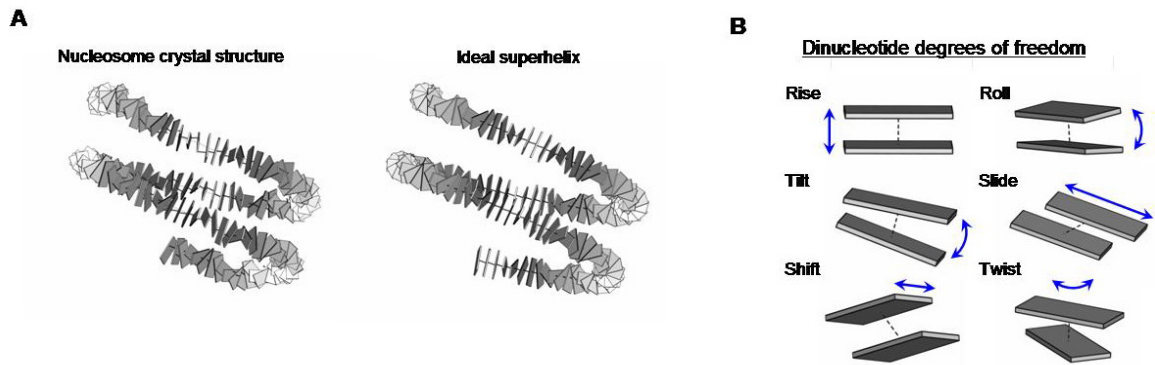

**Figure 6** a) DNA conformation from the crystal structure of the nucleosome core particle (PDB code 1kx5) (Richmond and Davey 2003) and the ideal superhelix. b) Conformation of a single dinucleotide (defined as two consecutive DNA base pairs in the 5'-3' direction) is described by six geometric degrees of freedom: three relative displacements (rise, shift, and slide) and three angles (twist, roll, and tilt). DNA base pairs are shown as rectangular blocks, and the direction of each displacement or rotation is indicated with arrows.

# III. Computational studies of chromatin structure

## 1. Using DNA elasticity theory to predict nucleosome formation energies

The availability of crystal and NMR structures of nucleosome core particles and other protein-DNA complexes makes it possible to predict free energies of nucleosome formation and nucleosome occupancy profiles *ab initio*, without resorting to high-throughput nucleosome positioning data sets described in Section II.1. When a nucleosome core particle is formed, a 147 bp long DNA molecule wraps around the surface of the histone octamer in ~1.65 turns of a left-handed superhelix (Richmond and Davey 2003). Because DNA wraps so tightly around the histone core (the length of nucleosomal DNA is comparable to the DNA persistence length), the free energy of bending DNA into a superhelical shape is strongly sequence-dependent: flexible sequences bend more easily than rigid DNA molecules such as poly(dA-dT) tracts. It is reasonable to assume that DNA bending depends mostly on base-stacking energies, so that the total free energy is given by the sum of individual dinucleotide contributions.

Typically, DNA base-stacking geometries are defined using three relative displacements (rise, shift and slide) and three rotation angles (twist, roll and tilt) for each dinucleotide formed by two adjacent base pairs (Figure 6b) (Olson *et al.* 1998). Together the six degrees of freedom completely specify the spatial position of base pair $i+1$ in the local coordinate frame attached to base pair $i$. Cartesian coordinates of an arbitrary DNA molecule can be used to construct a full set of relative dinucleotide geometries. Conversely, specifying a complete set of dinucleotide degrees of freedom is sufficient for reconstructing an arbitrary DNA conformation in global Cartesian coordinates (Lu *et al.* 1997a; Lu *et al.* 1997b).

As shown by Olson *et al*., these degrees of freedom can be used to derive an empirical model of DNA elastic energies (Olson *et al.* 1998). The model takes distributions of dinucleotide geometries observed in the ensemble of non-homologous protein-DNA structures as input (there are currently >100 such structures in the Protein Data Bank). These data are used to compute the mean and the covariance matrix for each degree of freedom (rise, shift, …) and for each dinucleotide type (AA, AC, …). Retaining the full covariance matrix allows the model to include correlations between different degrees of freedom. DNA elastic energy is described by an effective quadratic potential:

$$E_{el} = \frac{1}{2}\sum_{s=1}^{N}\left[\alpha^s - \left\langle\alpha^{n(s)}\right\rangle\right]^T F^{n(s)}\left[\alpha^s - \left\langle\alpha^{n(s)}\right\rangle\right],$$  (1)

where $\alpha^s$ is the six-component vector of dinucleotide degrees of freedom, the sum runs over $N = 146$ consecutive dinucleotides, and $\left\langle\alpha^n\right\rangle$ is the vector of average values for each degree of freedom for the dinucleotide or type $n$ at position $s$. $F^{n(s)}$ is a matrix of stiffness coefficients computed by inverting the covariance matrix $C^n$ for the dinucleotide of type $n$ at position $s$: $F^n = (C^n)^{-1}$, where

$$C_{ij}^n = \left\langle\left(\alpha_i^n - \left\langle\alpha_i^n\right\rangle\right)\left(\alpha_j^n - \left\langle\alpha_j^n\right\rangle\right)\right\rangle.$$  (2)

Note that the elastic energy model utilizes only the first and second moments of the empirical distributions of dinucleotide geometries – the available structural data is insufficient for retaining higher-order moments or for modeling more than two consecutive base pairs.

The elastic energy model described above was first adapted to predicting nucleosome positions in a 2007 paper by Tolstorukov *et al.* (Tolstorukov *et al.* 2007). It is known from the 1.9 Å resolution crystal structure of the nucleosome core particle (Richmond and Davey 2003) that dinucleotide positions at which nucleosomal DNA is kinked have large positive values of slide and large negative values of roll (solid lines with squares in Figure 7). The kinks mediate bending DNA into the nucleosomal shape and define "hot spots" at which high-affinity nucleosome positioning sequences have flexible dinucleotides.

The authors argue that lateral slide deformations observed at sites of local anisotropic bending define the superhelical trajectory of nucleosomal DNA. They show that slide accounts for over 90% of the overall pitch of nucleosomal DNA, with stepwise accumulation of net pitch in ~10-11 bp increments. The positive values of slide accompany DNA bending into the minor groove where roll is negative, and the negative values of slide appear with DNA bending into the major groove where roll is positive (Figure 7). Because the direction of slide is roughly parallel to the superhelical axis at the kink sites, the values of slide accumulate along the path of nucleosomal DNA. As a result, steps with alternating positive and negative slide (separated by ~5–6 bp) contribute cooperatively to the overall superhelical pitch.

While Tolstorukov *et al.* mostly focused on the contributions of roll and slide to the nucleosomal geometry, they also used DNA elastic energies to predict optimal positions of mononucleosomes reconstituted *in vitro* on four short DNA sequences and mapped by hydroxyl radical footprinting. Elastic energy was computed for each allowed position of the nucleosome core particle along each DNA segment, and energy minima were identified with predicted optimal positions. DNA geometry was taken from the high-resolution crystal structure (Richmond and Davey 2003), under the *ad hoc* assumption that sequence-dependent variations in DNA geometries and kink positions were not important for predicting nucleosomal energies. Regardless of this limitation the model showed reasonable predictive power, although it is difficult to judge how its performance would scale up in genome-wide calculations where multiple nucleosomes form simultaneously under steric constraints.

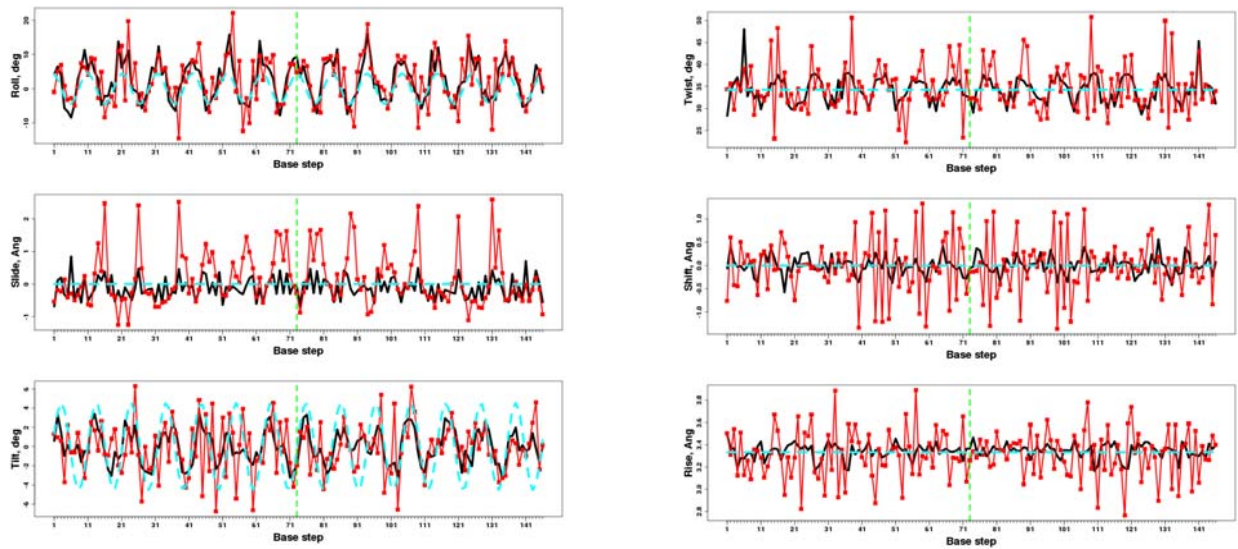

**Figure 7** Dinucleotide geometries from the crystal structure of the nucleosome core particle (solid red lines with squares, PDB code 1kx5) (Richmond and Davey 2003); from the minimum energy structure obtained by relaxing the 147 bp DNA segment with the nucleotide sequence from 1kx5 around the ideal superhelix as described in Section III.1 (solid black lines) (Morozov *et al.* 2009); and from the ideal superhelix (dashed cyan lines). Six dinucleotide degrees of freedom are shown: roll, slide, tilt, twist, shift, and rise. The two-fold nucleosome symmetry axis is shown as a dashed vertical line. Note that roll is negative when DNA bends into the minor groove (which faces the histone octamer) and positive when DNA bends into the major groove (with the minor groove facing away from the histone octamer).

Another approach utilizing the empirical DNA elastic potential was developed by Morozov *et al.* (Morozov *et al.* 2009). The main methodological difference with Tolstorukov *et al.* (Tolstorukov *et al.* 2007) is that the DNA geometry was allowed to be relaxed from its initial conformation (given *e.g.* by the ideal superhelix or by the nucleosome crystal structure). Thus the total energy of a nucleosomal DNA was given by a weighted sum of two quadratic potentials:

$$E = E_{el} + wE_{sh}, \tag{3}$$

where $E_{el}$ is the sequence-dependent DNA elastic potential and $E_{sh}$ is the non-specific histone-DNA interaction energy designed to penalize deviations of nucleosomal DNA from the ideal superhelix:

$$E_{sh} = \sum_{s=1}^{N} (\vec{r}_s - \vec{r}_s^{\,0})^2 \,, \tag{4}$$

where $\vec{r}_s$ and $\vec{r}_s^{\,0}$ are the nucleosomal DNA and the ideal superhelix radius-vectors to the origin of base pair $s$ in the global frame (the base pair origin is defined by its atomic coordinates (Lu *et al.* 1997b; Lu *et al.* 1997a)). While this term is an oversimplification of complex atomic interactions between histones and DNA, its quadratic form reduces minimization of the total energy $E$ to solving a system of linear equations. The final conformation of the DNA molecule is then the one that minimizes its total energy.

Using this approach the authors carried out the first *ab initio* prediction of DNA base step geometries in nucleosomal DNA. With only the 147 bp á-satellite sequence from 1kx5 as input, they predicted 6 dinucleotide degrees of freedom observed in the crystal structure with the average correlation coefficient of 0.46 (solid lines in Figure 7). In comparison, the average correlation coefficient between the ideal superhelical geometry and 1kx5 was 0.07. The model underpredicted the absolute values of slide in kink positions (although the overall correlation was above average at 0.54) and did not reproduce rapid alternate shifts between bps 31 and 111 (Figure 7). This behavior could be due to inaccuracies in the coefficients of the empirical elastic potential (derived from non-histone protein-DNA complexes), or due to the inherent limitations of the quadratic model (Eq. (3)).

The authors also predicted free energies of nucleosome formation for ~45 *in vitro* sequences (with the ~0.8 correlation coefficient), as well as nucleosome positions for six short sequences with nucleosomes mapped by hydroxyl radical footprinting (four from Tolstorukov *et al.* (Tolstorukov *et al.* 2007) and two more determined *de novo*). The authors found that DNA geometry relaxation helped with predicting nucleosome free energies but had surprisingly little effect on the accuracy of predicted nucleosome positions. Finally, the authors provided an exact numerical solution to the many-body problem of placing multiple nucleosomes onto longer stretches of DNA ((Durbin *et al.* 1998; Segal *et al.* 2006); see Section III.3 for details). Free energies of nucleosome formation at every DNA bp are translated by the algorithm into nucleosome probabilities and occupancies under the assumption of steric exclusion between nucleosome core particles of fixed size. The performance of the model on genome-wide data was found to be modest however, with 5' NDRs predicted to be nucleosome-enriched rather than depleted (Morozov *et al.* 2008).

Miele *et al.* have used a sequence-dependent DNA flexibility model to predict nucleosome occupancies in *S.cerevisiae* and *D.melanogaster* genomes. Their model employed the ideal superhelical geometry (with pitch and radius from 1kx5) and considered only angular (roll, twist and tilt) isotropic contributions to the superhelical curvature. In other words, DNA was modeled as an inextensible and unshearable elastic rod, and correlations between different degrees of freedom were neglected. The equilibrium values of the angular parameters and the stiffness coefficients were adopted from Anselmi *et al.* (Anselmi *et al.* 2000). These parameters are not based on protein-DNA structural data – rather, they were derived by energy calculations in the framework of the nearest-neighbor approximation (De Santis *et al.* 1986) and later refined to improve the correlation between calculated and experimental gel electrophoresis mobility of a large pool of synthetic and natural DNA molecules. The authors also took into account the entropic cost of the transition from free to superhelical DNA, and computed the free energy difference $\Delta F$ at every bp along the sequence.

In yeast, the free energy landscape was compared with the log2 ratio of hybridization values from Yuan *et al.* (Yuan *et al.* 2005) and Lee *et al.* (Lee *et al.* 2007). The authors observed significant correlation between their predictions and experimental data. In particular, the model was able to discriminate between sets of DNA fragments with the highest and lowest nucleosome occupancies (with the area under the receiver operating characteristic (ROC) curve of 0.72 for the data set from Yuan *et al.*). Furthermore, the model predicted 5′ NDRs. In the fly the model exhibited nucleosome occupancy depletion (*i.e.* higher values of $\Delta F$) in the vicinity of target sites of the Trithorax and Polycomb group proteins zeste and Ez/Psc. This depletion was also observed in the experimental profile of chromatin sensitivity to MNase obtained in a study of histone H3.3 replacement patterns in *D.melanogaster* (Mito *et al.* 2007).

DNA elastic parameters used by Miele *et al.* were also employed for nucleosome positioning predictions by the original authors of the force field, De Santis and colleagues (Anselmi *et al.* 2000; Anselmi *et al.* 2002; Scipioni *et al.* 2009). In addition to the DNA model based on roll, tilt and twist, the authors explicitly consider the conformational entropy contribution to the relative thermodynamic stability of the nucleosome core particle and also introduce an empirical term which depends on the curvature of free DNA. To justify the latter term, the authors observe that DNA elastic energies alone gives a satisfactory agreement for relatively straight DNA but exhibit large deviations for intrinsically curved DNA. The magnitude of these deviations shows a strong correlation with $\left\langle A_f^0 \right\rangle$, the average integral curvature of free DNA. On the basis of this observation the authors added an empirical contribution of $4.5 \left\langle A_f^0 \right\rangle^{3/2}$ to the model.

The De Santis *et al.* approach leads to accurate predictions of free energies of nucleosome formation ($r = 0.92$ on a set of ~100 mononucleosomal sequences collected from the literature) (Scipioni *et al.* 2009). The authors also compute free energy profiles for several genomic loci and compare them with experimental data from Yuan *et al.* (Yuan *et al.* 2005) and Kaplan *et al.* (Kaplan *et al.* 2009). Unfortunately, the authors do not provide overall measures of performance such as ROC curves, making it difficult to compare their approach with other methods.

Another model which considers DNA to be an unshearable elastic rod is due to Vaillant *et al.* (Vaillant *et al.* 2007). Although DNA is formally described by all three local angles (roll, tilt and twist), only the roll degree of freedom is sequence-dependent, while the equilibrium value of tilt is set to 0 and the equilibrium value of twist is set to $2\pi/10.5$ for all dinucleotides. Because the geometry of nucleosomal DNA is assumed to be described by an ideal superhelix, only the roll degree of freedom contributes to the overall elastic energy (cf. Eq. (1)). Equilibrium values of roll are based on a trinucleotide coding table from Goodsell and Dickerson (Goodsell and Dickerson 1994), who in turn adapt them from Satchwell *et al.* (Satchwell *et al.* 1986). The coding table uses a relatively small set of nucleosome positioning sequences to compute the fractional preference of each bp triplet to be outside or inside of the DNA wound around the histone octamer. "Outside" and "inside" refer to the position of the major groove with respect to the histone surface: for example, the GGC triplet has a 45% preference for locations on a bent double helix in which its major groove faces inward and is compressed by the curvature (corresponding to the region of the positive roll), whereas the AAA triplet has a 36% preference for the opposite orientation, with the major groove facing outward (corresponding to the region of the negative roll).

Despite being phrased in the language of DNA elastic energies, the Vaillant *et al.* model has more in common with statistical nucleosome positioning scores described in Section III.2 (Ioshikhes *et al.* 2006; Segal *et al.* 2006; Field *et al.* 2008; Kaplan *et al.* 2009) than with physical models. The roll degree of freedom is simply used to quantify relative orientational preferences of trinucleotides with respect to the helical twist, and could have been replaced by position-dependent scores without any explicit reference to DNA geometry.

## 2. Bioinformatics models of nucleosome sequence preferences

An alternative approach to predicting nucleosome positions is based on training bioinformatics models to discriminate various sequence features that differentiate nucleosomal and linker DNA. This approach requires extensive collections of nucleosomal sequences available from recent high-throughput nucleosome positioning experiments described in Section II.1 (we note in passing that two "first-generation" nucleosome positioning models, Ioshikes *et al.* (Ioshikhes *et al.* 2006) and Segal *et al.* (Segal *et al.* 2006) used much smaller, ~200 collections of nucleosome positioning sequences. These early approaches have been largely superseded by the more recent models trained on high-throughput data).

### 2.1. Segal *et al.* model based on dinucleotide distributions and 5-mer counts

One bioinformatics model that positions nucleosomes according to their sequence preferences was developed in (Segal *et al.* 2006; Field *et al.* 2008; Kaplan *et al.* 2009). This model assigns a nucleosome formation score to each 147 bp long DNA sequence. The score accounts for two major nucleosome positioning signals: the higher affinity of histone octamers for periodic distributions of certain dinucleotides due to anisotropic sequence-dependent DNA bending (rotational positioning), and longer motifs that function as nucleosome favoring or disfavoring signals, such as poly(dA-dT) tracts found mainly in the nucleosome-depleted regions (translational positioning). Similarly to DNA elastic energies, the bioinformatics score can be used to evaluate genome-wide nucleosome occupancies with steric exclusion, as described in Section III.3 below. The latest iteration of this model is trained on the *in vitro* high-throughput sequencing data set discussed in Section II.1.1.2, and is capable of predicting genome-wide nucleosome occupancies with a correlation coefficient of 0.89 for *in vitro* data and 0.75 for *in vivo* data (Kaplan *et al.* 2009).

As mentioned above, the Segal *et al.* model consists of two components, $P_N$ and $P_L$. The first component, $P_N$, captures position-dependent periodicity of dinucleotide distributions (Figure 5). In analogy with position-specific scoring matrices (Stormo and Fields 1998), the nucleosome-bound DNA sequences and their reverse complements are aligned around the dyad axis. The alignment is used to estimate the conditional dinucleotide distribution $P_{N,i}(S_i | S_{i-1})$ for each $i = 2 \ldots 147$ – the probability of observing the nucleotide $S_i$ at position $i$ given the nucleotide $S_{i-1}$ at position $i-1$. The first component of the model is then defined as:

$$P_N(S) = P_{N,1}(S_1) \prod_{i=2}^{147} P_{N,i}(S_i \mid S_{i-1}).$$ (5)

Thus to any DNA sequence $S$ of length 147 bp one can assign a probability using Eq. (5). Since this component captures nucleosomal dinucleotide sequence preferences, the probability is higher for sequences with ~10-11 bp phased patterns of A/T and G/C-rich dinucleotides (Figure 5).

The second component of the score, $P_L$, accounts for the difference in global, position-independent distributions of 5 bp long words in nucleosomal and linker sequences (5-mers are chosen here for computational reasons). To find out which of the 1024 sequences of length 5 is favored or disfavored by nucleosomes, probability $P_l$ is evaluated for each word of length 5 as the ratio between the frequency of that word in linkers and its frequency in the nucleosomal sequences. This ratio is normalized by the sum of all ratios across all words of length 5. So by definition sequences with higher probability will be less favorable for nucleosomes than sequences with lower probability. The position-independent component $P_L(S)$ for a given sequence $S$ of length 147 is then defined as a product over all instances of 5-mers in $S$:

$$P_L(S) = \prod_{i=5}^{147} P_l(S_i \mid S_{i-4}, \ldots, S_{i-1}).$$ (6)

By construction $P_L$ is higher for nucleosome-depleted sequences, in contrast with $P_N$ which is higher for nucleosome-bound sequences. Both of these components are used to predict nucleosome occupancies genome-wide (Segal *et al.* 2006; Kaplan *et al.* 2009; Segal and Widom 2009): $P_N$ captures rotational positioning while $P_L$ describes translational positioning. The log-ratio between the components

$$E(S) = -\log \frac{P_N(S)}{P_L(S)},$$ (7)

assigns a nucleosome formation "energy" to each sequence $S$ of length 147 bp. Segal *et al.* use the score given by Eq. (7) to formulate a thermodynamic model which predicts nucleosome occupancies and positions (see Section III.3).

### 2.2. Ioshikes *et al.* comparative genomics model

Another computational model that employed position-dependent periodicity of nucleosomal dinucleotide distributions was developed by Ioshikhes *et al.* (Ioshikhes *et al.* 2006). The authors used a periodic 139 bp dinucleotide AA/TT signal as an empirical definition of a nucleosome positioning sequence (NPS). The periodic pattern used in this work was compiled by Ioshikhes *et al.* in a previous study (Ioshikhes *et al.* 1996) where a collection of 204 nucleosomal DNA sequences was used to compute the AA/TT positional frequency distributions. The substantial noise in the data had been reduced by employing five different multiple sequence alignment techniques with subsequent averaging of the AA/TT positional frequency profiles.

In Ref. (Ioshikhes *et al.* 2006) the entire *Saccharomyces cerevisiae* genome was scanned for correlations with the NPS. To increase the method's resolution the authors utilized a comparative genomics approach, specifically, they calculated NPS correlation profiles for each gene between -1000 bp and +800 bp, relative to the +1 ATG start codon, in six sequenced *Saccharomyces* species. The authors argued that well-positioned nucleosomes should be conserved at orthologous locations in all related species. Thus averaging across orthologous locations in different *Saccharomyces* species would suppress noise and increase the sensitivity of the method compared to a single genome approach. The Ioshikhes *et al.* study revealed several nucleosome positioning features conserved across related yeast species. Specifically, most yeast promoters possess a nucleosome-depleted region, revealed by the most negative correlation between the underlying DNA sequence and the periodic NPS. Well-positioned +1 nucleosomes just downstream of the NDR (Figure 3) exhibit maximum positive correlation between the underlying DNA sequence and the NPS. The authors also pointed out a difference in correlation patterns for TATA-less and TATA-containing genes: TATA-less genes tend to have a more consistent NPS-NDR-NPS pattern.

## 2.3. Support vector machine for identifying the nucleosome formation potential

A discriminative approach to predicting nucleosome positions was developed in Ref. (Peckham *et al.* 2007). Specifically, Pekham *et al.* implemented a support vector machine (SVM) (Vapnik 1998) to distinguish nucleosome-forming and nucleosome-inhibiting DNA sequences. Because the SVM algorithm performs discriminative data classification in vector space, Pekham *et al.* converted each 50 bp long DNA sequence into a vector of k-mer frequencies, where k runs from 1 to 6. By means of this procedure every DNA sequence can be uniquely mapped into a 2772-dimensional vector space. The SVM algorithm finds a hyperplane in this vector space which separates two groups of training data in such a way that the distance from the hyperplane to the nearest data point is maximized.

The training data used by Pekham *et al.* was taken from a low-resolution microarray study of Yuan *et al.* (Yuan *et al.* 2005) (see Section II.1.1.1) where nucleosome occupancies were measured for yeast chromosome III. The training set consisted of 1000 nucleosome-forming (50 bp long) DNA sequences with the highest hybridization scores and 1000 nucleosome-inhibiting (50 bp long) DNA sequences with the lowest hybridization scores. An SVM trained on this data set can classify DNA segments as nucleosome-forming or nucleosome-inhibiting after mapping them into vector space.

The authors employed the trained SVM to estimate the nucleosome formation potential of a collection of ~200 nucleosomal DNA sequences studied in Segal *et al.* (Segal *et al.* 2006). The nucleosome formation potential correlated with the dinucleotide AA/TT/AT periodic signal found in Ref. (Segal *et al.* 2006). The authors argued that G/C and A/T content of a sequence is the strongest predictive factor in determining the nucleosome formation potential: G/C-rich sequences favor nucleosome formation while A/T-rich sequences disfavor it. They also pointed out that A-tracts tend to prevent nucleosome formation. Based on how well the SVM discriminated between nucleosome-bound and nucleosome-free genomic sequences, the authors concluded that only ~50% of all nucleosomes are positioned by intrinsic DNA sequence signals.

### 2.4. Wavelet-based approach to discriminating nucleosome-bound from nucleosome-free sequences

Yuan and Liu proposed an alternative computational model for predicting nucleosome positions which is based on a wavelet transform (Yuan and Liu 2008). Similarly to Ref. (Peckham *et al.* 2007), the authors developed a classification algorithm that distinguishes nucleosome-bound sequences from linker DNA. Each DNA sequence $S$ of length 131 bp is decomposed into 16 numerical vectors of length 130 bp, where each vector describes one of 16 dinucleotides. The $i$-th component of this vector is 1 if the corresponding dinucleotide is found at location $i$ in sequence $S$, and 0 otherwise. The moving average with a 3 bp window is applied to the vectors resulting in sequences of length 128 bp (the length $128 = 2^7$ is motivated by the discrete wavelet transform). Thus each of the 16 vectors holds information about the corresponding dinucleotide frequency in $S$.

The method used in Ref. (Yuan and Liu 2008) is based on a wavelet transform which allows to detect periodic patterns in the signal over multiple scales (Mallat 1999). In the wavelet analysis the signal is decomposed into orthogonal components corresponding to different frequency bands. The advantage of this approach over Fourier spectral analysis is that the wavelet transformation captures not only the frequency component but also the position of this component in the signal. The coefficients of the wavelet transform characterize periodic patterns embedded in the dinucleotide frequency signal.

Thus, if a certain frequency is associated with a putative nucleosome positioning sequence, it can be detected by comparing the contributions (defined by the coefficients of the wavelet transform) from nucleosomal vs. linker DNA sequences. The probability for a DNA sequence $S$ to be a nucleosomal sequence is then defined through a logistic regression model with the covariates given by the coefficients of the discrete wavelet transform. The model was trained on previously identified nucleosome and linker sequences (Yuan *et al.* 2005; Segal *et al.* 2006). In contrast with the Pekham *et al.* model (Peckham *et al.* 2007), this approach only accounts for dinucleotide frequencies. The authors conclude that sequence information is highly predictive of local nucleosome enrichment or depletion, whereas predictions of the exact nucleosome positions are only moderately accurate, suggesting the importance of other regulatory factors in fine-tuning nucleosome positions.

## 3. Statistical physics of one-dimensional liquids and the nucleosome positioning problem

Recent high-resolution maps of nucleosome locations in eukaryotic genomes (Lee *et al.* 2007; Mavrich *et al.* 2008a; Mavrich *et al.* 2008b; Schones *et al.* 2008; Shivaswamy *et al.* 2008; Valouev *et al.* 2008) reveal that nucleosomes are arranged in quasi-periodic arrays, covering ~70-80% of genomic DNA. Furthermore, regions upstream of open reading frames are typically depleted of nucleosomes. Nucleosome-depleted promoter regions are flanked on both sides by the so-called ±1 nucleosomes (Jiang and Pugh 2009). It is believed (Mavrich *et al.* 2008a) that these nucleosomes act as boundaries, inducing periodic oscillations in the neighboring nucleosomes (Figure 3). These oscillations in nucleosome occupancy are caused by steric exclusion, and can be explained with a simple statistical model of a one-dimensional (1D) liquid. In this model the nucleosomes are treated as a uniform 1D liquid of 147 bp hard rods. The simplest approach of this kind assumes that nucleosomes have no intrinsic sequence specificity and are positioned solely by steric exclusion (Kornberg and Stryer 1988). To induce periodic oscillations in the nucleosome occupancy, a boundary constraint is introduced at some point along the DNA. The boundary may be due to the sequence-specific ±1 nucleosomes or to DNA-bound non-histone proteins.

Nucleosome reconstitution *in vitro* involves stepwise dialysis from concentrated salt solutions (when DNA molecules have low affinity for the histone octamer due to electrostatic screening) down to physiological salt concentrations (Shrader and Crothers 1989; Shrader and Crothers 1990; Lowary and Widom 1997). This is expected to lead to an equilibrium distribution of nucleosomes which can be described by equilibrium statistical mechanics. Nucleosome positions *in vivo* are influenced by the presence of chromatin remodelers and transcription factors and many other factors which in principle may perturb the system out of equilibrium. However, Kaplan *et al.* found that correlation between *in vitro* and *in vivo* occupancy profiles is quite high, more than 70% on average (Kaplan *et al.* 2009). Thus both *in vivo* and *in vitro* chromatin can be described using equilibrium methods.

In the Kornberg and Stryer model nucleosomal arrays can be regarded as an ensemble of two kinds of objects, nucleosomes of size $a$ bp and DNA linkers with the mean length of $L$ bp. The probability of selecting a nucleosome from the ensemble of a single nucleosome and $L$ linkers of unit length is $p = (1+L)^{-1}$, and the probability of selecting a linker is $1-p$. A stretch of DNA of length $x$ can contain any number of nucleosomes between 0 and $x/a$. Let $M \equiv \mathrm{int}(x/a)$ be the maximum number of nucleosomes that can still fit into the DNA domain of length $x$. The partition function for this domain is then defined as a binomial sum over all possible configurations of nucleosomes and linkers:

$$\Xi(x) = \sum_{n=0}^{M} \binom{n+x-an}{n} p^n (1-p)^{x-an} .$$
(8)

Kornberg and Stryer used this model to demonstrate how nucleosomes without intrinsic sequence preferences can nonetheless be organized into periodic arrays (Kornberg and Stryer 1988). Eq. (8) is used to calculate the probability of any configuration of nucleosomes and linkers on a DNA molecule of arbitrary length $N$. The probability $P(i)$ that site $i$ is occupied by a linker (*i.e.* is nucleosome-free) is derived by noting that the linker site $i$ divides the DNA molecule into two regions with lengths $i-1$ and $N-i$ and the linker site itself:

$$P(i) = \Xi(i-1) \frac{1-p}{\Xi(N)} \Xi(N-i) .$$
(9)

The normalization factor $\Xi^{-1}(N)$ corresponds to all possible configurations of nucleosomes and linkers over the entire DNA molecule.

The linker probability profiles (Eq. (9)) evaluated for various values of $L$ are shown in Figure 8. We observe a regular spacing of nucleosomes near the boundaries with the oscillation period of $a+L$. The oscillations decay with the distance from the boundary. Thus the boundary constraint alone leads to an array of regularly spaced nucleosomes at non-random locations. However, the significance of such boundaries in genomes and their molecular nature are still unclear. The nucleosome-positioning boundaries may be defined by sequence-specific DNA-bound proteins (Fedor *et al.* 1988; Roth *et al.* 1990; Pazin *et al.* 1997), or by rigid DNA sequences such as Poly(dA-dT) tracts that tend to exclude nucleosomes (Iyer and Struhl 1995; Suter *et al.* 2000; Anderson and Widom 2001; Bao *et al.* 2006; Lee *et al.* 2007; Mavrich *et al.* 2008a; Segal and Widom 2009).

In the Kornberg and Stryer statistical model nucleosome positions are dictated solely by steric constraints. Such a model does not account for histone-DNA interactions which should introduce a certain amount of sequence specificity into nucleosome positioning on DNA. However, it is straightforward to extend the model to the sequence-specific case and thus predict nucleosome occupancies for histone-DNA interactions of arbitrary magnitude.

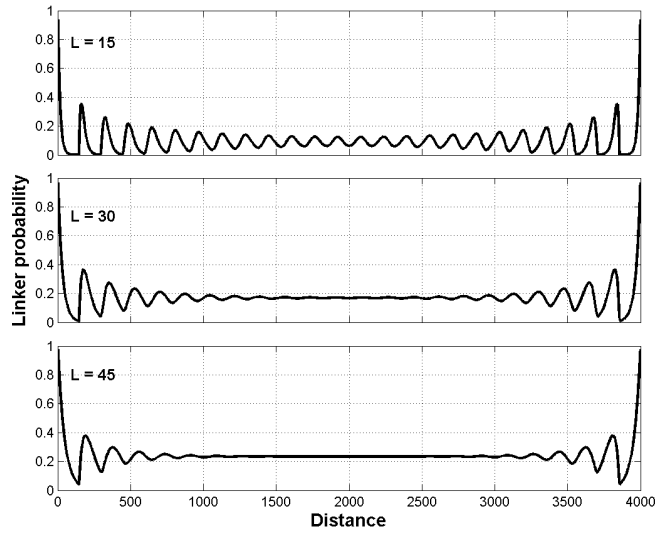

**Figure 8** Statistical positioning of nucleosomes by the boundaries. Linker probabilities computed using Eq. (9) with $N = 4000$, $a = 147$ and three different average linker lengths $L$ are shown. The period of the oscillations depends on the linker length as $a + L$.

For DNA sequence $S$ and a set of $k$ nucleosomes at positions $n_1, n_2, \ldots, n_k$ one defines a statistical weight function as a product of the corresponding Boltzmann factors:

$$W[S, n] = \prod_{i=1}^{k} \tau \exp\left\{ -\beta E(S_{n_i, n_i + 146}) \right\},\qquad(10)$$

where $n$ denotes the specific configuration of nucleosomes, $\beta$ is the inverse temperature, $S_{n_i, n_i + 146}$ is the part of sequence $S$ covered by the nucleosome $n_i$, and $\tau$ is the apparent nucleosome concentration (Field *et al.* 2008). The nucleosome positions $n_1, n_2, \ldots, n_k$ are chosen in such a way that no two nucleosomes overlap. To find the probability of a given configuration $n$ of nucleosomes on $S$ the statistical weight is divided by the partition function which is a sum over all allowed nucleosome configurations: $P(W[S,n]) = W[S,n] / \sum_{n'} W[S,n']$.

The number of allowed configurations for long DNA sequences is exponentially large. Nevertheless, the probability of starting a nucleosome at every genomic bp can be efficiently computed using the dynamic programming method (Durbin *et al.* 1998). The idea behind the method is similar to that used in deriving Eq. (9): the nucleosome that starts at base pair $i$ divides DNA sequence $S$ of length $N$ into three subsequences $S_{1,i-1}$, $S_{i,i+146}$, and $S_{i+147,N}$. In the forward step one finds the partial partition function $F_{i-1}$ which is the sum of the nucleosome weight functions over all possible configurations on $S_{1,i-1}$. The partition function $R_{i+147}$ corresponding to the sequence $S_{i+147,N}$ is found in the backward step. The forward and backward steps are further explained in Figure 9. The probability that a nucleosome starts at base pair $i$ of DNA sequence $S$ is then given by

$$P_i = \frac{F_{i-1} \tau \exp\{-E(S_{i,i+146})\} R_{i+147}}{R_1}, \qquad (11)$$

where $R_1 = F_N$ in the denominator is the complete partition function. The corresponding nucleosome occupancy $O_i$ is found as the probability that a given base pair $i$ is covered by any nucleosome, and is defined as the sum of probabilities from $P_{i-146}$ to $P_i$.

Although steric exclusion is taken into account in the probability profile $P_i$, this profile is derived for a specific value of the chemical potential $\mu$ which enters the model through the nucleosome concentration $\tau = \exp(\beta\mu)$ in Eq. (11). Variation of this parameter leads to different probability and occupancy profiles due to the competition between the nucleosome binding energy and the excluded volume interaction. The resulting change in occupancy can be quite substantial so it is reasonable to ask about the behavior of the model with respect to $\mu$. In principle, this question can be addressed by re-running the dynamic programming algorithm for various values of $\tau$. However, Schwab *et al.* (Schwab *et al.* 2008) used a more physical approach which employs statistical mechanics of a 1D liquid of hard rods in an arbitrary external field (Percus 1976).

The system is assumed to be in a grand-canonical ensemble. The first component of the statistical model is the potential energy $V_i$ of a nucleosome positioned at base pair $i$ and related to the probability $P_i$ from Eq. (11) through $V_i = -k_B T_0 \log P_i$. The second component of the model is the interparticle potential responsible for steric hindrance. This model is solvable in a sense that the nucleosome distribution can be computed exactly for various values of the chemical potential $\mu$ and the temperature $T_0$ which specifies the characteristic energy scale (Schwab *et al.* 2008).

**Forward solution:**

$$F_0 = F_1 = \ldots = F_{a-1} = 1$$
$$F_i = F_{i-1} + F_{i-a}\tau \exp\{-\beta E_{i-a+1}\}, \, i = a \ldots N$$

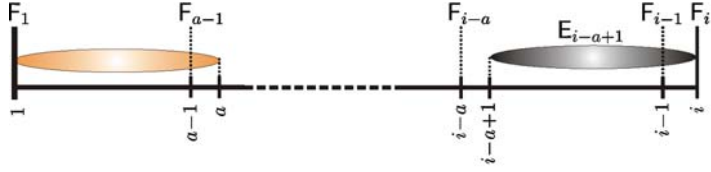

**Backward solution:**

$$R_{N-a+2} = \ldots = R_{N+1} = 1$$
$$R_i = R_{i+1} + R_{i+a}\tau \exp\{-\beta E_{i+a-1}\}, \, i = 1 \ldots N-a+1$$

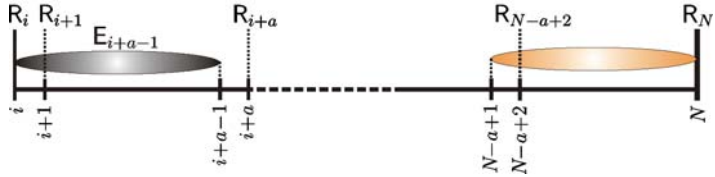

**Figure 9** Schematic representation of the forward and backward steps (Eq. (11)). The positions of nucleosomes of size $a$ that are close to the boundary (light ovals) are restricted: no nucleosome can end at positions $1 \ldots a-1$ or start at positions $N-a+2 \ldots N$. This leads to a set of boundary conditions for forward and backward partial partition functions. The forward partition function $F_i$ corresponds to all possible configurations of nucleosomes on the interval from $1$ to $i$. If $i \geq a$ there exist two possibilities: 1) position $i$ is empty; this corresponds to the first term $F_{i-1}$ which indicates that the nucleosome configuration is unchanged from $i-1$ to $i$; 2) a nucleosome which started at position $i-a+1$ ends at position $i$ (dark ovals); this case is accounted for by the second term in the recursive equation which is the product of the nucleosome Boltzmann factor and the partition function $F_{i-a}$ for the interval $[1 \ldots i-a]$. The backward partition function $R_i$ is evaluated in the opposite direction and describes nucleosome configurations on the interval from $i$ to $N$.

Both the energy profile $V_i$ and the excluded volume interaction define the final disposition of nucleosomes on DNA. Due to the competition between these two terms a small change in the chemical potential or the strength of nucleosome binding can lead to repositioning of some of the nucleosomes (Schwab *et al.* 2008), as shown schematically in Figure 10. Whereas positions of more stable nucleosomes remain unchanged, positions of less stable nucleosomes shift with $\mu$. The mean number of nucleosomes $\langle N \rangle$ as a function of the chemical potential is shown in Figure 10 for two values of temperature $T/T_0$. The observed behavior, which resembles a first-order phase transition, is a consequence of the fact that at the transition point the free energy of a given nucleosome arrangement is degenerate with respect to $N$. It has been demonstrated in Ref. (Schwab *et al.* 2008) that changes in nucleosome occupancy are localized to certain regions, and that the genomic locations of these regions correlate with known positions of transcription factor binding sites.

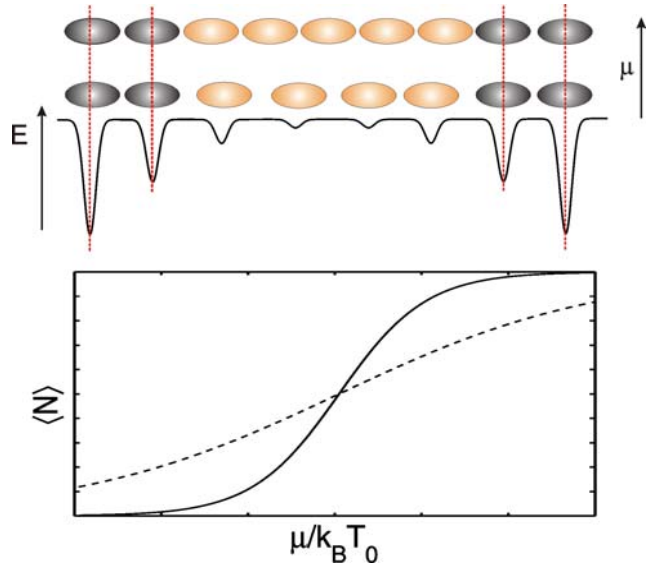

**Figure 10** Top, a nucleosome configuration is shown for two different values of the chemical potential. Stable nucleosomes (dark ovals) remain at their positions while unstable nucleosomes (light ovals) are shifted and another nucleosome appears as ì is increased. Bottom, the average number of nucleosomes on an arbitrary DNA segment is shown schematically as a function of ì for two different temperatures. Dashed line corresponds to the higher temperature.

In the statistical models discussed here it is assumed that apart from steric overlap nucleosome positions are independent and are dictated solely by intrinsic sequence preferences. In general this approximation may not hold. The internucleosomal potential can originate, for instance, from the higher-order structure of the chromatin fiber. Fiber formation causes linker lengths to be quantized (Kato *et al.* 2003; Cohanim *et al.* 2006; Wang *et al.* 2008) because relative spatial positions of adjacent nucleosomes depend on the length of their linker DNA (Widom 1992; Schalch *et al.* 2005). Another possible source of the internucleosomal potential is electrostatic interactions between spatially close nucleosomes (Luger *et al.* 1997; Dorigo *et al.* 2004; Chodaparambil *et al.* 2007). Quantized linker lengths can be described with an effective nearest-neighbor internucleosomal potential. The thermodynamic model of Eq. (11) can be modified to account for nearest-neighbor interactions by introducing an additional term that assigns different statistical weights to different linker lengths (Lubliner and Segal 2009).

## 4. Hidden Markov Models for predicting nucleosome occupancies

Log-intensity profiles obtained from microarray-based nucleosome positioning experiments are often analyzed using Hidden Markov Models (HMMs) (Rabiner 1989). Although HMMs were originally developed for analyzing sequential stochastic signals such as noisy time-series data, they have recently found many uses in bioinformatics. In the context of chromatin structure prediction the use of HMMs was pioneered by Yuan *et al.* (Yuan *et al.* 2005) and later adopted with modifications by Lee *et al.* (Lee *et al.* 2007) and Zawadzki *et al.* (Zawadzki *et al.* 2009). In each of these applications, hybridization values from the tiled array (*i.e.* log2 ratios of the nucleosomal DNA to the total genomic DNA for each microarray probe) were used as input to the HMM, which then predicted the probability of a nucleosome to start at every genomic bp as well as nucleosome occupancy (defined as the probability that a given bp is covered by any nucleosome). Thus HMMs are used to calculate nucleosome occupancies directly from the log-intensity data, in contrast with the approaches that first predict a sequence-specific nucleosome free energy profile and then use dynamic programming to infer nucleosome positions.

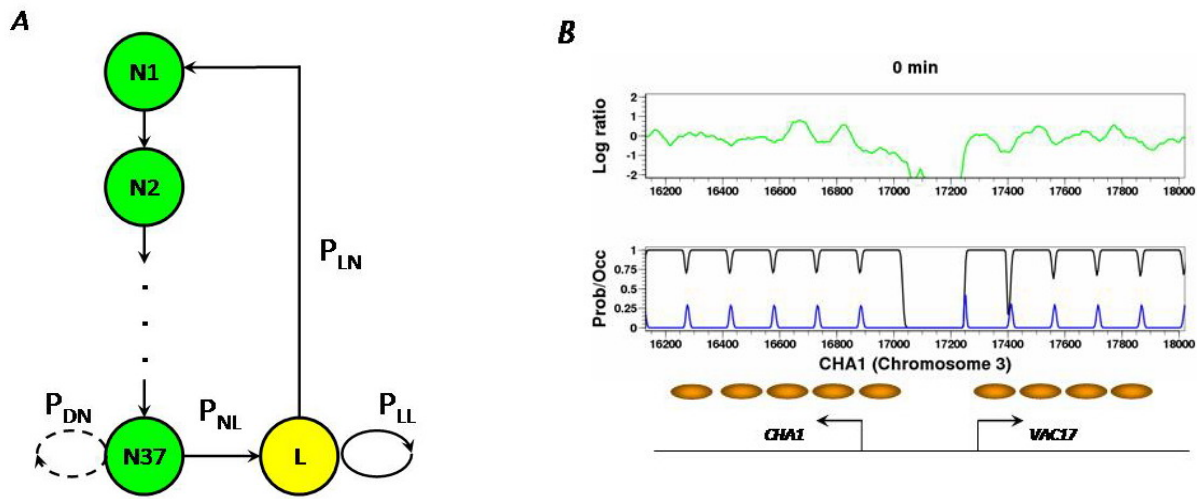

**Figure 11** a) The transition matrix of the Hidden Markov Model (HMM) used for predicting nucleosome probabilities and occupancies from microarray log-intensity profiles. Nodes $N_1$ through $N_{37}$ represent consecutive nucleosomal states (green circles), while $L$ is the linker state (yellow circle). Once a new nucleosome is started 36 subsequent nodes are placed with probability 1. From $N_{37}$ a transition is made to the linker state with probability $P_{NL}$, or else the nucleosome is extended indefinitely to create "delocalized" nucleosomal states (Yuan *et al.* 2005; Lee *et al.* 2007). The linker state can also be extended indefinitely with probability $P_{LL}$ or else the transition is made back to the first nucleosome state $N_1$. Note that $P_{NL} + P_{DN} = 1$, $P_{LL} + P_{LN} = 1$. b) Nucleosome positions prior to glucose addition at the *CHA1* promoter predicted by the HMM. Top, log ratio of nucleosomal DNA to genomic DNA as determined by separate hybridizations to Affymetrix tiling arrays is plotted as a function of genomic position. Increasing values represent increasing protection from MNase digestion. Middle, nucleosome positions predicted by the HMM (Zawadzki *et al.* 2009). Black trace represents predicted nucleosome occupancy, from unoccupied (0) to fully occupied (1). Blue trace represents the probability of starting a nucleosome at a given bp. Bottom, previously mapped *in vivo* nucleosome positions (Moreira and Holmberg 1998) are shown as dark orange ovals.

HMMs are defined by the transition matrix between hidden states which are probabilistically assigned to each consecutive probe on the tiling microarray. For Affymetrix arrays with probes tiled at every 4 bp, 37 nucleosomal node and 1 linker node need to be defined, resulting in a 148 bp long nucleosome core particle (Zawadzki *et al.* 2009). As shown in Figure 11a, from the linker state a transition can be made to another linker state with probability $P_{LL}$, or to the first nucleosome state with probability $P_{LN} = 1 - P_{LL}$. However, once a new nucleosome is started all subsequent nodes are placed with probability 1 until the next linker state is reached.

The model used in Zawadzki *et al.* employs a mixture of two Gaussians to represent both the nucleosome state and the linker state (earlier models used a single Gaussian). Thus the final set of fitting parameters includes means and widths of four one-dimensional Gaussians, two independent mixture coefficients, and 38 initial probabilities for the nucleosome and linker states. To reduce the number of fitting parameters, 37 nucleosome states were forced to share the same mixture of two Gaussians. All fitting parameters were found by maximum likelihood using standard methods (Rabiner 1989; Durbin *et al.* 1998). Note that the probability of starting a new nucleosome ($P_{LN}$ in Figure 11a), fit by maximum likelihood in Refs. (Yuan *et al.* 2005) and (Lee *et al.* 2007), was adjusted manually in Ref. (Zawadzki *et al.* 2009) in order to achieve a pre-defined average nucleosome occupancy of ~80%. The manual adjustment reflects lack of control over the zero intensity baseline, which may shift depending on the relative amounts of DNA used in the nucleosomal and control samples.

In Refs. (Lee *et al.* 2007) and (Zawadzki *et al.* 2009) which employ high-density Affymetrix arrays, HMM parameters were fit for only a relatively small set of genomic loci (*e.g.* promoters and coding sequences of 19 genes whose expression was unaffected by adding glucose to the medium in Ref. (Zawadzki *et al.* 2009)), and averaged. The resulting model was then run genome-wide with fixed parameters. As a typical example, an HMM prediction for the *CHA1* locus in which nucleosomes were previously mapped using low-throughput methods (Moreira and Holmberg 1998) is shown in Figure 11b.

Another important distinction between the three HMM approaches is that Refs. (Yuan *et al.* 2005) and (Lee *et al.* 2007) allow "delocalized" nucleosomes of arbitrary length (the modified topology of the transition matrix is shown in Figure 11a with dashed lines), whereas Ref. (Zawadzki *et al.* 2009) postulates only canonical, 148 bp nucleosomes and thus interprets longer stretches with high log-intensity ratios as shifted but overlapping nucleosome positions in distinct subpopulations of cells.

## IV. Summary and Conclusions

It is amazing to see by how much the field of high-throughput mapping of nucleosome positions has advanced over just a few years. Whereas in 2005 the first tiling microarray covered only yeast chromosome III with 20 bp resolution (Yuan *et al.* 2005), today's high-throughput sequencing platforms yield hundreds of millions of nucleosome positions. Microarray and high-throughput sequencing technologies are to some extent complementary: interrogating a genomic region with tiled microarrays yields nucleosome occupancy in that region, but the approach does not easily scale up to higher eukaryotes with much longer genomes. Thus the best currently available microarray data for bulk nucleosomes in *D.melanogaster* (122.6 Mbp genome length) is at 36 bp resolution (Mavrich *et al.* 2008b), while in *H.sapiens* (3300 Mbp genome length) only select regions have been interrogated (Dennis *et al.* 2007; Ozsolak *et al.* 2007).

Unlike microarrays, high-throughput sequencing is not restricted to a particular genomic region: nucleosome sequence reads can come from any genomic locus. However, even latest parallel sequencing data sets do not provide enough read coverage to measure relative nucleosome occupancies in longer genomes, although the data can still be used to infer common nucleosome positioning motifs (such as the periodic distribution of AA/AT/TA/TT dinucleotides shown in Figure 5) and to study nucleosome organization in the vicinity of coding sequences and TF binding sites. To circumvent the problem of low sequence read coverage, several recent studies of chromatin structure in human and fly chose to focus on the nucleosomes which incorporate the H2A.Z histone variant or have acetylated/methylated histone tails (Barski *et al.* 2007; Mavrich *et al.* 2008b). These partial maps yield comprehensive genome-wide coverage for a given nucleosomal subspecies but in general cannot be extrapolated to account for bulk nucleosomes.

All collections of nucleosome sequences published to date have utilized single-end reads: only one end of ~150 bp mononucleosome cores is sequenced and mapped onto the reference genome. The other end of the nucleosome core particle has to be inferred by adding 147 bp to the starting position of the sequence read. This procedure assumes that MNase treatment liberates nucleosome core particles precisely, without leaving undigested DNA at the nucleosome termini or digesting nucleosome-covered DNA. A measure of MNase cutting precision is provided by plotting a correlation function between starting coordinates of sequence reads mapped onto the Watson (W) and Crick (C) DNA strands (Figure 4). Although there is a distinct maximum at 0 bp lag (which corresponds to the 147 bp separation between starting positions of sequence reads mapped onto the W and C strands), the width of the peak clearly shows that the majority of mononucleosome cores are not isolated with bp precision. The question of how precisely nucleosome core particles are located with respect to sequence read coordinates can be further addressed using paired-end reads (in which, as its name suggests, both ends of the DNA molecule are sequenced and mapped, so that its length is known exactly).

Another potential issue in nucleosome positioning studies is MNase sequence specificity. It is unlikely that MNase actively repositions nucleosomes, because experiments done with and without nucleosome cross-linking yield similar patterns of nucleosome organization and similar sequence determinants of nucleosome positioning (Kaplan *et al.* 2009). However, the question of whether MNase binding specificity can bias which nucleosomes get sequenced (by preferentially isolating mononucleosome cores flanked by "good" MNase binding sites) is not fully resolved in the literature. While some authors argue that MNase binding specificity is negligible (Ozsolak *et al.* 2007; Field *et al.* 2008), others explain observed A/T enrichment in regions flanking nucleosome core particles as a consequence of preferential cleavage by micrococcal nuclease (Johnson *et al.* 2006; Valouev *et al.* 2008).

High-throughput nucleosome positioning studies helped establish a canonical picture of nucleosome organization in genic and intergenic regions. As shown in Figure 3, most yeast genes are flanked by 5' and 3' NDRs which help arrange genic nucleosomes into quasi-periodic arrays. Although most authors agree that intrinsic sequence preferences play a certain role in establishing *in vivo* nucleosome positions, there is a range of opinions as to how important this role is.

One series of studies argues on the basis of strong correlation between sequence-based bioinformatics models and *in vivo* occupancy profiles that most nucleosomes in living cells are positioned by sequence (Segal *et al.* 2006; Field *et al.* 2008; Kaplan *et al.* 2009). A competing view, the so-called "barrier model" of nucleosome positioning, emphasizes that regular nucleosomal arrays can be created simply by steric exclusion (Mavrich *et al.* 2008a; Zhang *et al.* 2009). The ends of such arrays may be defined by nucleosome-excluding sequence elements, by "anchoring" nucleosomes optimized for binding affinity and/or stabilized by interactions with other proteins and protein complexes, or by DNA-bound TFs. According to this view, most nucleosomes need not be sequence-specific. It is interesting to note that there are significant differences between *in vitro* and *in vivo* nucleosome occupancy profiles: the 5' NDR is significantly less pronounced *in vitro* and there are no nucleosome-size oscillations over ORFs (Figure 3). Thus nucleosomes are not intrinsically ordered with respect to the TSS and borders of *in vivo* arrays are shaped by (yet unknown) external factors, *e.g.* by interactions with the components of transcription initiation machinery (Zhang *et al.* 2009). Surprisingly, there is no difference between *in vitro* and *in vivo* 3' NDRs, which thus appear to be established mainly through nucleosome-disfavoring sequences.

In this review we have deliberately not focused on the nucleosome occupancy in the vicinity of TF binding sites. It is harder to see the general picture here because the relationship between nucleosomes and TFs is much more varied, and because precise genomic locations of TF binding sites are not always known. Nevertheless, it appears that nucleosomes are intrinsically depleted over some types of TF binding sites, whereas for other factors (such as Abf1 and Reb1 in yeast) there are marked differences between *in vivo* and *in vitro* nucleosome occupancies (Kaplan *et al.* 2009). In accordance with this view, chromatin appears surprisingly stable with respect to environmental (Shivaswamy *et al.* 2008; Zawadzki *et al.* 2009) and genetic (Whitehouse *et al.* 2007) perturbations. Overall, there are only minor changes in nucleosome positions and numbers, indicating that chromatin is largely "pre-set" for transcriptional response.

Computational approaches to predicting nucleosome occupancies can be based on either physics or bioinformatics. Physical models of nucleosome formation energies employ DNA elasticity theory (in some cases augmented with additional terms that take into account conformational entropy, intrinsic DNA curvature, etc.) to compute the sequence-dependent free energy of bending the 147 bp long DNA into a nucleosomal superhelix. Free energies computed at every bp along the DNA sequence can then be used as input to the dynamic programming algorithm (Durbin *et al.* 1998; Segal *et al.* 2006; Morozov *et al.* 2009) which solves the many-body problem of positioning multiple nucleosomes on DNA without steric overlap (although most physics-based studies do not attempt to derive nucleosome occupancies from free energy predictions). Unlike bioinformatics approaches, physical models do not utilize training sets of nucleosome positioning sequences, which could lead to biased predictions if all sequences came from a particular genome or if the sequence set was insufficiently large. However, elastic potentials typically depend on empirical coefficients such as equilibrium values of DNA geometric parameters. Such coefficients have to be estimated from structural data or molecular mechanics simulations, making the models dependent on the quality of the estimates.

In contrast to the bioinformatics approaches designed to search for DNA sequence signals that discriminate between nucleosome-enriched and nucleosome-depleted regions, physical models can explain observed sequence patterns in terms of elastic energies associated with DNA bending and the corresponding geometries of the nucleosomal DNA. If the DNA conformation is allowed to relax, such models are even capable of predicting the minimum energy DNA conformation for comparison with crystal structures (Morozov *et al.* 2009). It appears that physical models can predict free energies and *in vitro* positions of single nucleosomes reconstituted on artificial and natural sequences reasonably well (Tolstorukov *et al.* 2007; Morozov *et al.* 2009; Scipioni *et al.* 2009). Their genome-wide accuracy is less clear: some papers do not make genome-wide predictions at all (Tolstorukov *et al.* 2007; Morozov *et al.* 2009) while others provide limited comparisons which do not include latest data sets or bioinformatics models (Miele *et al.* 2008; Scipioni *et al.* 2009). It would be of great interest to test physical models against each other and against their bioinformatics counterparts using latest high-throughput parallel sequencing data sets and a uniform set of performance metrics.

Bioinformatics approaches employ a wide range of statistical techniques (including support vector machines, wavelet analysis, and Markov models) to assign nucleosome positioning scores. Although a rigorous comparison between all published models is not available, it is fair to say that the latest generation of bioinformatics models can be used to discriminate nucleosome-enriched from nucleosome-depleted regions with high accuracy and to predict general features of nucleosome organization such as 5' and 3' NDRs. An interesting hybrid approach has been developed by Eran Segal and co-workers (Segal *et al.* 2006; Field *et al.* 2008; Kaplan *et al.* 2009). As discussed in detail in Section III.2.1, the authors compute a nucleosome positioning score by first making an alignment of a large number of nucleosomal sequences obtained from a high-throughput parallel sequencing run. Next they define a log-score based on: a) the dinucleotide distribution at each position in the nucleosomal site; b) the position-independent distribution of 5 bp words inside and outside nucleosomes. The bioinformatics scores at each genomic position are then treated as "energies" and used to predict genome-wide nucleosome occupancy profiles by solving the many-body problem of positioning multiple nucleosomes on the genomic DNA (Section III.3). The latest iteration of this model is based on the *in vitro* nucleosome positioning data set and is capable of predicting the *in vitro* occupancy profile with the correlation coefficient of 0.89 (Kaplan *et al.* 2009).

We expect that a more uniform view of the factors responsible for *in vivo* nucleosome positioning will emerge in the near future. Comparisons between various models and a careful analysis of observed nucleosome sequence features will help establish the limits of applicability and the relative strengths and weaknesses of alternative modeling approaches. Future computational predictions of nucleosome positions should be equally applicable to genomic sequences from multiple organisms and to synthetic DNA, and will establish the relative importance of intrinsic nucleosome sequence preferences in maintaining and regulating *in vivo* chromatin.

## Acknowledgements

The authors are grateful to George Locke for assistance with high-throughput sequencing data, and to Karl Zawadzki for carefully proofreading the manuscript. A.V.M. and D.T. were supported by a grant from the National Institutes of Health (R01 HG-004708). A.V.M. was also supported by an Alfred P. Sloan Research Fellowship.